\pdfoutput=1
 \documentclass[aps,prl,twocolumn,showpacs,superscriptaddress,preprintnumbers]{revtex4-1}  

\usepackage{amsmath}
\usepackage{amssymb}
\usepackage{epsfig}
\usepackage{graphicx}
\usepackage{hyperref}
\usepackage{tabu}
\usepackage{boldline}
\usepackage{slashed}
\usepackage{multirow}
\usepackage{diagbox}

\usepackage{color}
\usepackage[normal]{subfigure}
\usepackage[table]{xcolor}
\usepackage{enumitem}
\usepackage[utf8]{inputenc}
\usepackage{colortbl}
\definecolor{nicered}{rgb}{0.7,0.1,0.1}
\definecolor{nicegreen}{rgb}{0.1,0.5,0.1}
\definecolor{red}{rgb}{1.0, 0, 0}

\usepackage{lineno}

\definecolor{LightCyan}{rgb}{0.88,1,1}
\definecolor{piggypink}{rgb}{0.99, 0.87, 0.9}
\definecolor{applegreen}{rgb}{0.55, 0.71, 0.0}
\definecolor{darkpastelgreen}{rgb}{0.01, 0.75, 0.24}
\definecolor{green-yellow}{rgb}{0.68, 1.0, 0.18}

\newcommand{\beq}{\begin{equation}}
\newcommand{\eeq}{\end{equation}}
\newcommand{\beqa}{\begin{eqnarray}}
\newcommand{\eeqa}{\end{eqnarray}}

\newcommand{\units}[1]{{\, \rm #1}}

\graphicspath{{figs/}}

\begin{document}


\title{SENSEI: First Direct-Detection Constraints \\ on sub-GeV Dark Matter from a Surface Run 
}

\author{The SENSEI Collaboration: Michael Crisler}
\email{mike@fnal.gov}
\affiliation{\normalsize\it 
Fermi National Accelerator Laboratory, PO Box 500, Batavia IL, 60510}

 \author{Rouven Essig}
\email{rouven.essig@stonybrook.edu}
\affiliation{\normalsize\it 
C.N.~Yang Institute for Theoretical Physics, Stony Brook University, Stony Brook, NY 11794}

 \author{Juan Estrada}
\email{estrada@fnal.gov}
\affiliation{\normalsize\it 
Fermi National Accelerator Laboratory, PO Box 500, Batavia IL, 60510}

\author{Guillermo Fernandez}
\email{fmoroni.guillermo@gmail.com}
\affiliation{\normalsize\it 
Fermi National Accelerator Laboratory, PO Box 500, Batavia IL, 60510}

 \author{Javier Tiffenberg}
\email{javiert@fnal.gov}
\affiliation{\normalsize\it 
Fermi National Accelerator Laboratory, PO Box 500, Batavia IL, 60510}

 \author{Miguel Sofo Haro}
\email{miguelsofoharo@gmail.com}
\affiliation{\normalsize\it 
Fermi National Accelerator Laboratory, PO Box 500, Batavia IL, 60510}
\affiliation{Centro At\'omico Bariloche, CNEA/CONICET/IB, Bariloche, Argentina}

 \author{Tomer Volansky}
\email{tomerv@post.tau.ac.il}
\affiliation{\normalsize\it 
Raymond and Beverly Sackler School of Physics and Astronomy, \\
 Tel-Aviv University, Tel-Aviv 69978, Israel}
 \affiliation{\normalsize\it
 School of Natural Sciences, The Institute for Advanced Study, Princeton, NJ 08540, USA}

 \author{Tien-Tien Yu}
\email{tien-tien.yu@cern.ch}
\affiliation{\normalsize\it 
Theoretical Physics Department, CERN, CH-1211 Geneva 23, Switzerland}
\affiliation{\normalsize\it 
Department of Physics and Institute of Theoretical Science, University of Oregon, Eugene, Oregon 97403}

\preprint{YITP-SB-18-4, CERN-TH-2018-070}
\begin{abstract}
  \noindent
The Sub-Electron-Noise Skipper CCD Experimental Instrument (SENSEI) uses the recently developed Skipper-CCD technology 
to search for electron recoils from the interaction of sub-GeV dark matter particles with electrons in silicon.  
We report first results from a prototype SENSEI detector, which
collected 0.019~gram-days of commissioning data above ground at Fermi National Accelerator Laboratory.  
These commissioning data are sufficient to set new direct-detection constraints for dark matter particles with masses between $\sim$500~keV 
and 4~MeV.  
Moreover, since these data were taken on the surface, they disfavor 
previously allowed strongly interacting dark matter particles with masses between $\sim$500~keV and a few hundred MeV.   
We discuss the implications of these data for several dark matter candidates, including one model proposed to explain 
the anomalously large 21-cm signal observed by the EDGES Collaboration.  
SENSEI is the first experiment dedicated to the search for electron recoils from dark matter, and these 
results demonstrate the power of the Skipper-CCD technology for dark matter searches. 
 \end{abstract}

\maketitle

 \section{Introduction}  

Identifying the nature of dark matter (DM) is one of the most important tasks of particle physics today, and direct-detection experiments play an 
essential role in this endeavor.  The search for DM particles with masses $\lesssim 1$~GeV represents a new experimental 
frontier~\cite{Battaglieri:2017aum}.  Traditional direct-detection searches, which are sensitive to recoil energy generated from DM scattering off 
of nuclei, typically have very little sensitivity to sub-GeV mass DM.  Indeed, the best current bounds are limited to very large cross sections 
below 1~GeV and are absent below $\sim 120$~MeV~\cite{Angloher:2017sxg,Petricca:2017zdp,Agnese:2015nto}. 
As suggested in~\cite{Essig:2011nj}, improved sensitivity to DM masses well below the GeV scale is possible by searching for signals 
induced by inelastic processes, for which a DM particle is able to deposit much more energy compared to the elastic scattering off of nuclei.  
In particular, one of the most promising avenues is to search for one or a few ionized electrons that are released due to 
DM particles interacting with electrons in the detector.

Background-free searches for single or few-electron events are experimentally challenging.  
Sensitivity to such events has been demonstrated using two-phase time projection chambers with noble liquid 
targets,  using data from XENON10~\cite{Essig:2012yx,Essig:2017kqs,Angle:2011th}, 
XENON100~\cite{Essig:2017kqs,Aprile:2016wwo}, and DarkSide-50~\cite{Agnes:2018oej}.
Significant progress can be made by utilizing solid-state detectors, which exhibit much lower thresholds due to their low, ${\mathcal O}$(eV), 
band gaps.  
Recently, silicon Charge-Coupled Devices (CCDs) with a special ``Skipper'' readout stage~\cite{Tiffenberg:2017aac} 
and high-voltage cryogenic silicon detectors with transition edge sensor readout~\cite{Romani:2017iwi} have demonstrated 
single-electron sensitivity.
The $\sim$1.1~eV band gap of silicon 
allows for  a DM mass threshold that is an order of magnitude lower than that achieved in noble liquid detectors, 
and permits significantly larger DM-electron scattering rates~\cite{Essig:2011nj,Essig:2015cda,Essig:2017kqs,Lee:2015qva,Graham:2012su}. 

The Sub-Electron-Noise Skipper CCD Experimental Instrument (SENSEI) is designed to utilize the Skipper-CCD technology 
demonstrated in~\cite{Tiffenberg:2017aac} to search for electron recoils from sub-GeV DM. 
While the ultimate goal of the SENSEI Collaboration is to build a 100-gram detector consisting of multiple Skipper-CCDs, 
a prototype detector is currently operating $\sim$100~m underground near the MINOS experiment 
at Fermi National Accelerator Laboratory (FNAL).  
This prototype was first tested on the surface, collecting 0.019-gram-days of data (before analysis cuts).  

In this letter, we present the first constraints on sub-GeV DM derived from SENSEI commissioning data. 
We exclude novel parameter space 
for DM masses below $\sim$4~MeV, above which the XENON10 constraint from~\cite{Essig:2012yx,Essig:2017kqs} dominates. 
Furthermore, operating on the surface allows a search for DM that strongly interacts with the visible sector.  
Such DM does not penetrate the Earth, and detectors placed deep underground, such as the noble-liquid detectors mentioned above, 
have no sensitivity.  
Despite large cosmic-ray backgrounds, this region can be easily probed by a detector on the surface 
with a small amount of data.  
The SENSEI data thus also place novel constraints on DM particles with masses of several hundred MeV.

\section{The SENSEI Prototype Detector }

We use a single Skipper-CCD of active area 1.086~cm $\times$ 1.872~cm with an 
initial active mass of 0.0947 grams of silicon fabricated parasitically in a production run 
for astronomical CCDs. 
The Skipper-CCD was packaged in a light-tight copper housing that was cooled to an estimated 130~K to reduce the dark current on 
the sensor and to reduce the emission of infrared photons from black-body 
radiation~\footnote{The CCD temperature is measured with an RTD placed some distance away, resulting in 
an uncertainty of a few degrees.}. 
The sensor was read by a modified Monsoon electronics system described in~\cite{Tiffenberg:2017aac}.  

We analyze here a small amount of commissioning data, taken on May 11, 2017, at the Silicon Detector Facility (SiDet) 
at FNAL.   SiDet has an elevation of $\sim$220~m above sea level and a roof 
consisting of about 7.6~cm of concrete, 2~mm of aluminum, and 1~cm of wood.  The thickness of the light-tight
copper housing in which the sensor was placed is 3~mm.  

\begin{figure}[t]
\begin{center}
\includegraphics[width=0.5\textwidth]{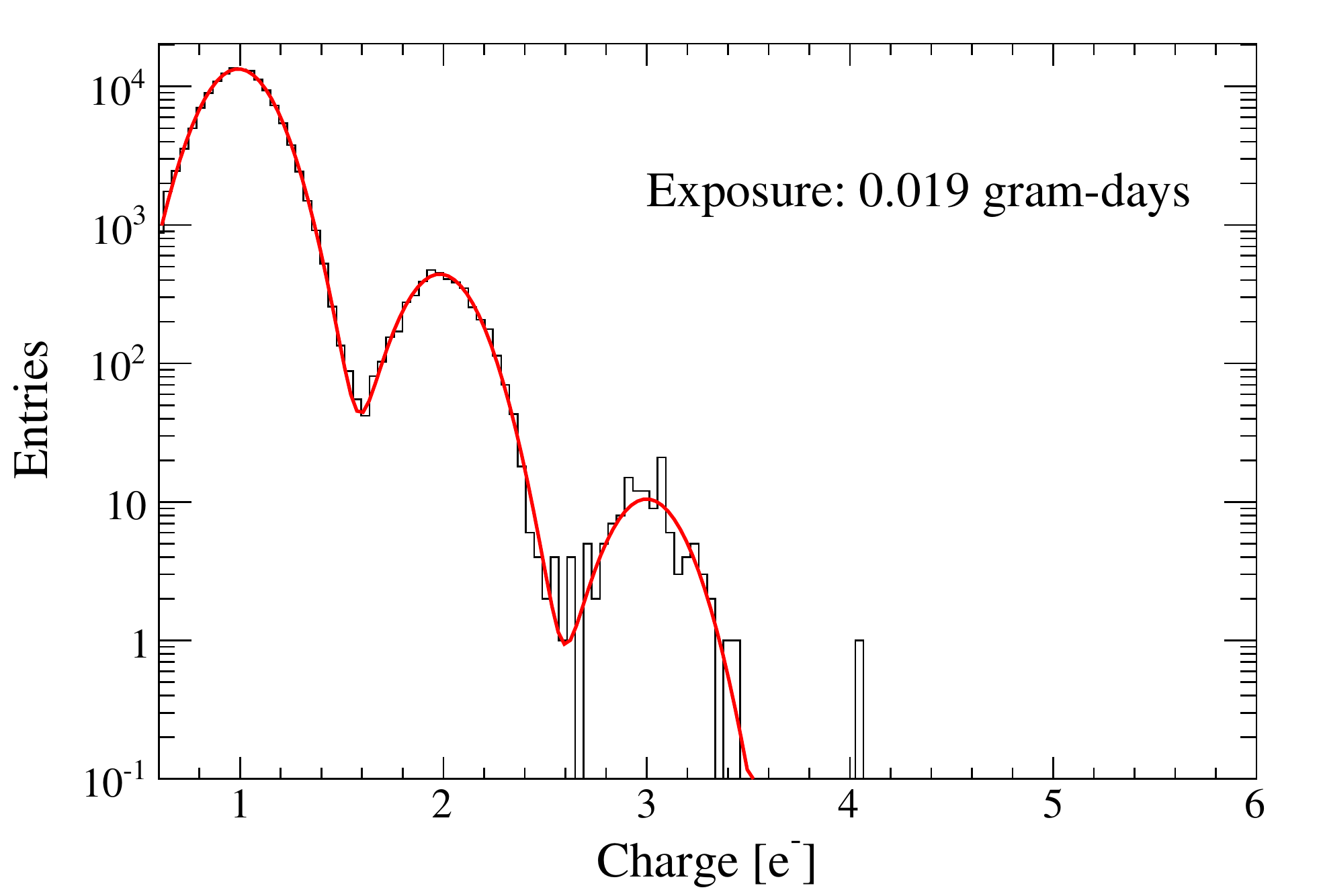}
\caption{Recorded spectrum after selection cuts for the 0.019 gram-days of commissioning data.  Gaussian fits to the peaks show there are 140,302, 4676, 131, and 1 event(s) with $1$, $2$, $3$, and $4$ electrons, respectively.  No events are seen for $5-100$~electrons. 
The gaussian width of the peaks are $\sim$0.14$e^-$.  
}
\label{fig:spectrum}
\end{center}
\end{figure}%

The Skipper-CCD is divided into four equal-size quadrants, each of which is 
read continuously, independently, and in parallel for 427~minutes.  
The single-sample noise of the CCD varied slightly from quadrant to quadrant. 
One of the quadrants had unusually high single-sample readout noise due to a charge transfer inefficiency problem in the readout stage, 
and all data from it were immediately discarded,
leaving an active mass of 0.071~grams.  
The other three had a single-sample readout noise of $\sim\! 4\units{e^-}$; 
taking 800 samples reduced the noise to $\sim$0.14$\units{e^-}$. 
It took $t_{\rm pix}\simeq 19.5495\times 10^{-3}$~s to take 800 samples of a single pixel. 

Each quadrant consists of 624 rows of 362 pixels.  
Each pixel has an area of $15 \mu\textrm{m} \times 15\mu\textrm{m}$, a thickness of 200$\mu$m, and a mass 
of $1.0476\times 10^{-7}$~gram.  
The total data is saved in 18 ``images'' for each quadrant, with an image 
consisting of 200 rows of 362 pixels.  
Before data taking, the CCD is cleaned, removing any excess charge on all pixels.   
The first 624 recorded rows in each quadrant then have pixels whose exposure grows linearly from $t_{\rm pix}$ to 
$624\times 362\times t_{\rm pix} = 73.6$~min, with the pixels in the following ($18\times 200 - 624=2976$) rows having a uniform exposure of 73.6 mins each.  
Note that due to the continuous readout, each row subsequent to the first 624 rows has a uniform exposure, corresponding to the 
time it takes to move a charge packet across the entire array of the CCD quadrant. 
The total exposure for the 3 ``active'' quadrants is 0.019 gram-days.

\section{Data Selection}
\label{sec:data}

The prototype Skipper-CCD has a measured dark current of $\sim 1.14\units{e^-/pixel/day}$. 
This leads to a large number of 1-, 2-, and 3-electron events.  
In this paper, we do not attempt to analyze the dark current in detail or to remove any background events.  
Instead, we place conservative limits by assuming that all events are from DM.

\begin{table}[t]
\begin{center}
\begin{tabular}{|l||c|c|c|c|c|}
\hline
 \diagbox{Cuts}{ $N_{e, \rm min}$ } & 1 & 2 & 3 & 4& 5 \\ \hline \hline
1.~DM within a single pixel & 1 & 0.62 & 0.48 & 0.41 & 0.37 \\ 
2.~Nearest Neighbor  & 0.8 & 0.8 & 0.8 & 0.8 & 0.8 \\ 
3.~Noise & 0.88 & 0.88 & 0.88 & 0.88 & 0.88 \\ 
4.~Bleeding & 0.95  & 0.95 & 0.95& 0.95& 0.95\\ \hline
Total & 0.67 &  0.41 & 0.32 & 0.27 & 0.24 \\ \hline\hline
Number of events& 140,302 & 4,676 & 131 &1 &0\\ \hline
\end{tabular}
\caption{Efficiencies for the data selection cuts for events with 1 to 5 electrons. The bottom row lists the number of observed events after cuts.
\vspace{-6mm}
} 
\label{tab:eff}
\end{center}
\end{table}%

\begin{figure*}[t]
\begin{center}
\includegraphics[width=0.32\textwidth]{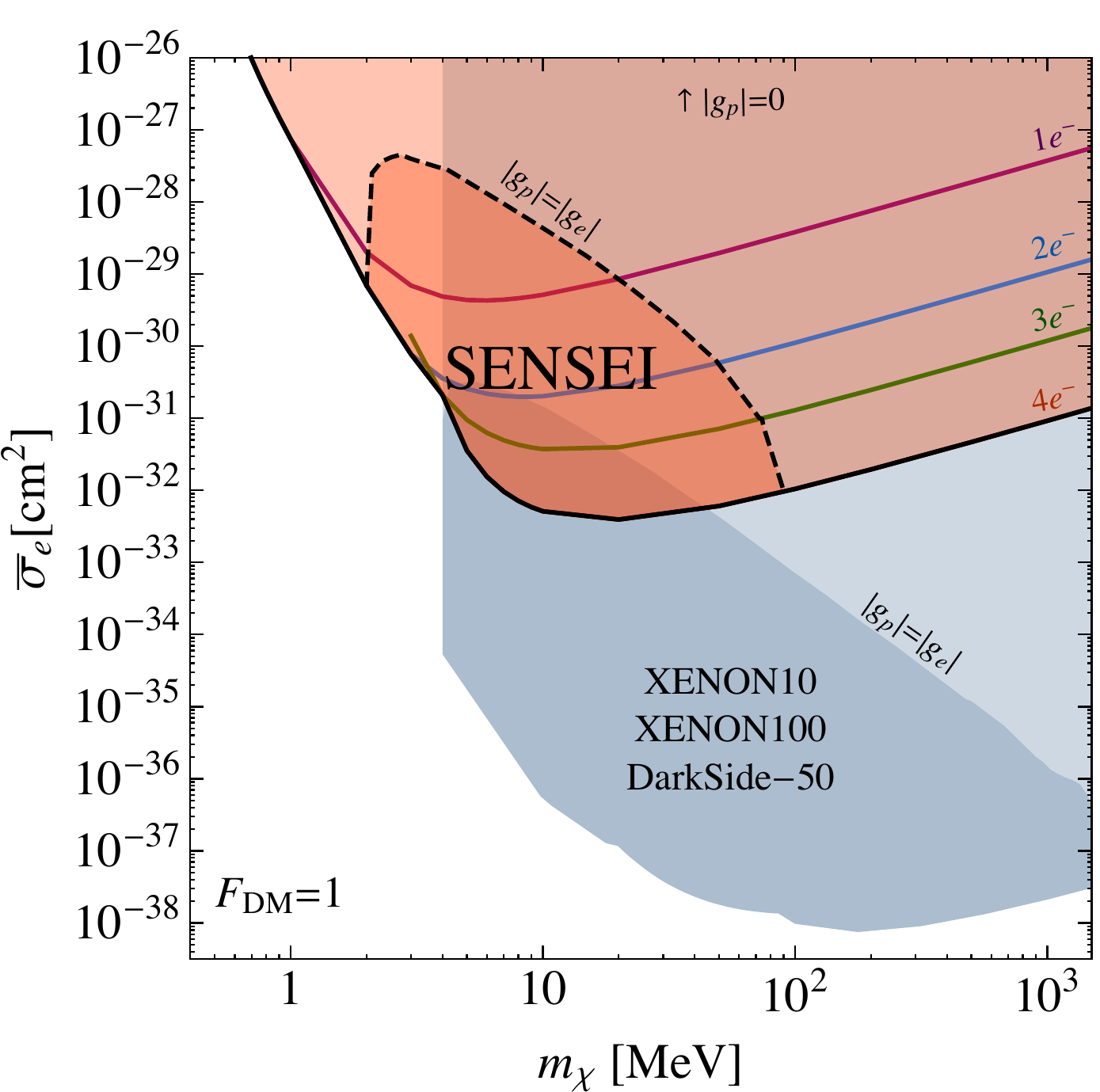}
\includegraphics[width=0.32\textwidth]{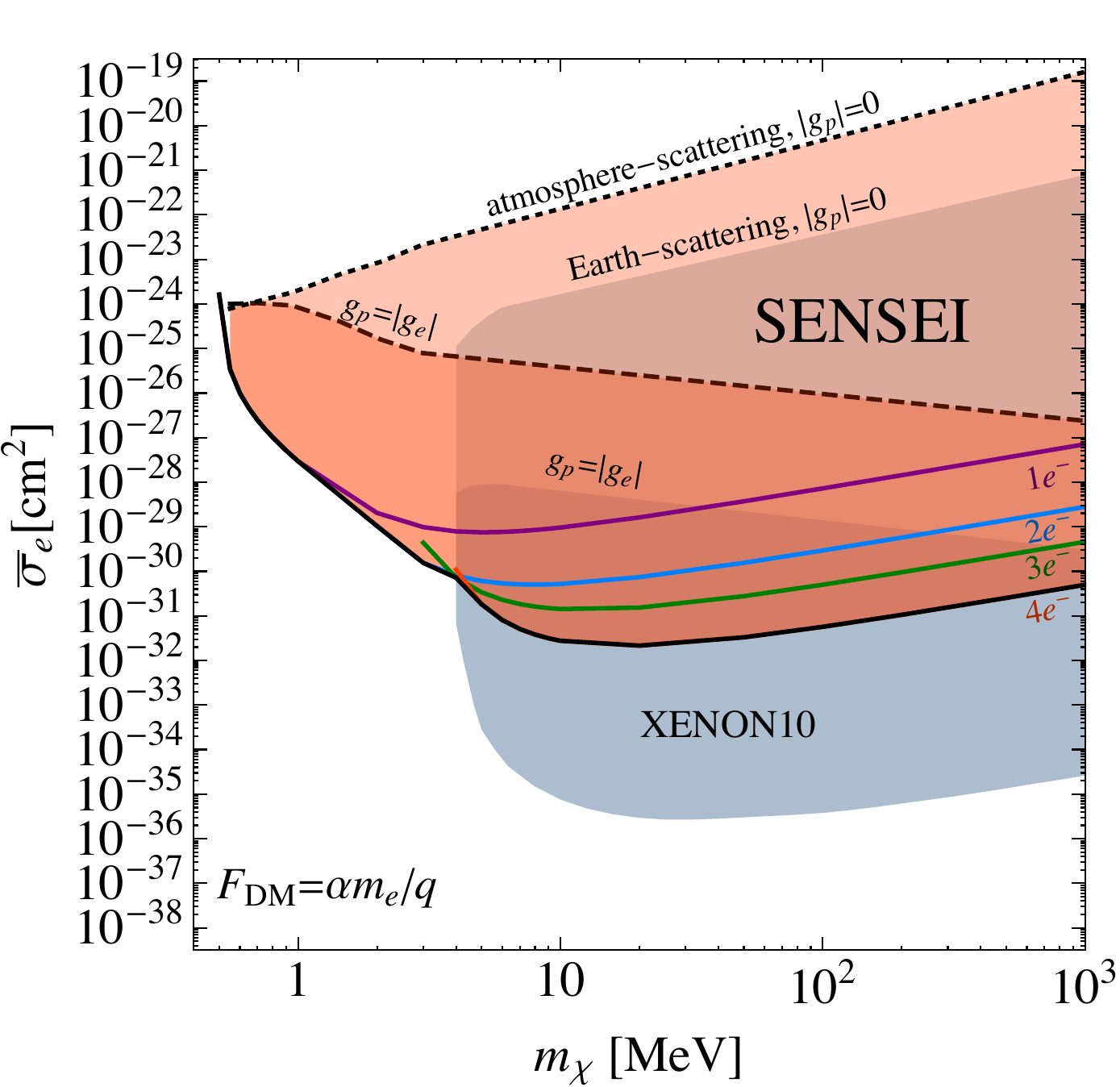}
\includegraphics[width=0.32\textwidth]{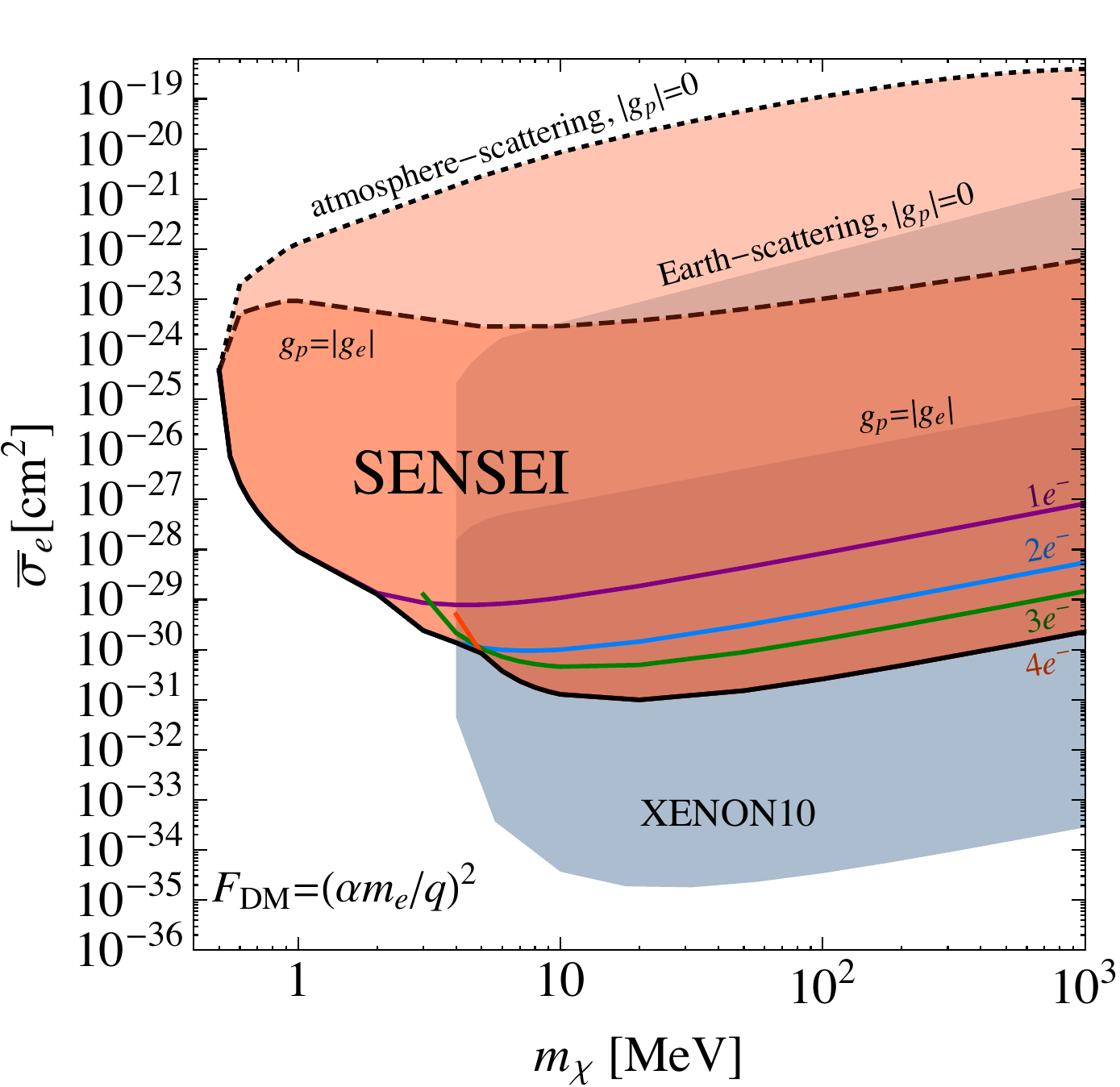}
\caption{
The 90\% C.L.~constraints on the DM-electron scattering cross-sections, $\overline{\sigma}_e$, 
as a function of DM mass, $m_\chi$, from a commissioning run above ground at FNAL using the SENSEI prototype detector.  
We show different DM form factors, $F_{\rm DM}(q)=1$, $\alpha m_e/q$, and $(\alpha m_e/q)^2$. 
The purple, blue, green, and red lines correspond to the constraints from the 1-, 2-, 3-, or 4-electron bin, respectively. 
The black line is the minimum of these. 
The blue shaded regions are the current constraints from XENON10, XENON100, and 
DarkSide-50. 
For large cross sections, the DM is stopped in the Earth's crust (atmosphere) and does not reach the noble-liquid (SENSEI prototype) 
detectors: the dark-shaded regions (labelled $|g_p|=|g_e|$) show order-of-magnitude estimates 
of the excluded parameter regions assuming the interaction between DM and ordinary matter is mediated 
by a heavy dark photon (left), an electric dipole moment (middle), or an ultralight dark photon (right).  
The light-shaded regions (labelled $g_p=0$) are order-of-magnitude estimates of 
the  90\% C.L. excluded parameter regions assuming a mediator that 
couples only to electrons.  
The terrestrial effects shown here are order-of-magnitude estimates only, and more detailed calculations will appear in~\cite{Emken:2018}.
}
\label{fig:DMescattering}
\end{center}
\end{figure*}

After data collection, we implemented several standard quality cuts for CCD-based 
detectors~\cite{Aguilar-Arevalo:2016khx, Aguilar-Arevalo:2016ndq} as well as cuts specific to our analysis, whose selection efficiencies 
are listed in Table~\ref{tab:eff} for electron bins $1-5$:
\begin{itemize}[leftmargin=*]\addtolength{\itemsep}{-0.6\baselineskip}
\item {\bf Single-pixel events}. 
To simplify our analysis, we select only pixels whose neighboring pixels are empty. 
\begin{itemize}[leftmargin=3mm]
\item {\bf DM within a single pixel}. 
While a DM event consists of one or more electrons that are created initially in a single pixel, 
and spread with uniform probability along the height of the pixel, 
the electrons can drift apart as they diffuse to the surface, 
allowing some electrons to diffuse to a neighboring pixel. 
The lateral drift distance is described by a gaussian distribution with a standard deviation proportional to the transit time from the interaction point to the surface of the CCD~\cite{Holland:2003}. 
The diffusion parameter can be measured directly from tracks produced by 
atmospheric muons~\cite{Aguilar-Arevalo:2016ndq}. 
For our CCD, the diffusion parameter is proportional 
to 0.002/$\mu$m times the interaction depth. 
The probability for the DM event to be contained in a single pixel is unity for 1-electron events, and drops to 0.166 for 100-electron events. 
\item {\bf Nearest Neighbor}.  
We remove all pixels that are next to an occupied pixel in the data; this cut also removes all tracks and clusters. 
\end{itemize}
\item {\bf Noise}.  We veto images in which the readout noise is 30\% larger than the expected readout noise as inferred from an 
over-scan region in which virtual (non-existent) pixels are read.  
\item {\bf Bleeding}.  At low temperatures the electron mobility may be impeded, implying a small probability that an electron can get stuck 
in a given pixel for several downward shifts.  
If an event, such as a cosmic ray, produces a large number of electrons in some 
pixel(s), then pixels with several electrons may be found upstream in the image.  
We mask 10 pixels upstream of any pixel containing more than 100 electrons.  
\end{itemize}

In what follows we bin the data, after the above selection cuts, according to the number of electrons per pixel, and derive constraints 
for each bin separately.  
The spectrum after cuts is shown in Fig.~\ref{fig:spectrum}, together with gaussian fits to the first three bins.   
We use the bins with $1-100$ electrons in our analysis.

\section{Analysis and Results}
We  calculate the DM recoil spectrum for several models, deriving 
constraints both on DM-electron scattering and on bosonic 
DM being absorbed by an electron~\cite{An:2014twa,Bloch:2016sjj,Hochberg:2016sqx,Hochberg:2016ajh}.   
For the scattering case, we use the calculations and conventions from~\cite{Essig:2011nj,Essig:2015cda}, assuming a local DM density 
$\rho_{\rm DM} =0.4\units{GeV/cm^3}$~\cite{Salucci:2010qr}.  
We present our results in the $\overline\sigma_e$ versus $m_\chi$ parameter space for various DM form-factors, $F_{\rm DM}(q)$, where
$m_\chi$ is the DM mass and $\overline\sigma_e$
is the cross section for DM to scatter off a free electron with the momentum 
transfer fixed to its typical value, $q=\alpha m_e$, where $\alpha$ is the fine-structure constant and $m_e$ is the electron mass.  
$F_{\rm DM}(q)$ parameterizes the model-dependent momentum dependence of the DM interaction: a ``heavy'' mediator with 
mass~$\gg\alpha m_e$ has $F_{\rm DM}(q)=1$; an ``ultralight'' mediator 
with mass~$\ll\alpha m_e$ has $F_{\rm DM}(q)=(\alpha m_e/q)^2$; 
and an electric-dipole-moment interaction with the Standard-Model photon produces $F_{\rm DM}(q)\simeq\alpha m_e/q$.  

For bosonic DM, we will consider that the DM is a dark photon, denoted $A'$, with mass $m_{A'}$, that is stable on the lifetime of the Universe.  
We follow the calculations and conventions in~\cite{Bloch:2016sjj}, and present results in the $\epsilon$ versus $m_{A'}$ 
parameter space, where $\epsilon$ is the parameter that characterizes the strength of the kinetic mixing between the $A'$ and the 
photon.  

For each model, we calculate conservative 95\% confidence level upper limits using Poisson statistics and assuming that all 
observed electrons in a given bin are DM events. 
We compare the resulting limit from each bin with the predicted number of DM events (for a given value of $\overline\sigma_e$ 
or $\epsilon$), after correcting for the efficiencies.  

\begin{figure}[t]
\begin{center}
\includegraphics[width=0.42\textwidth]{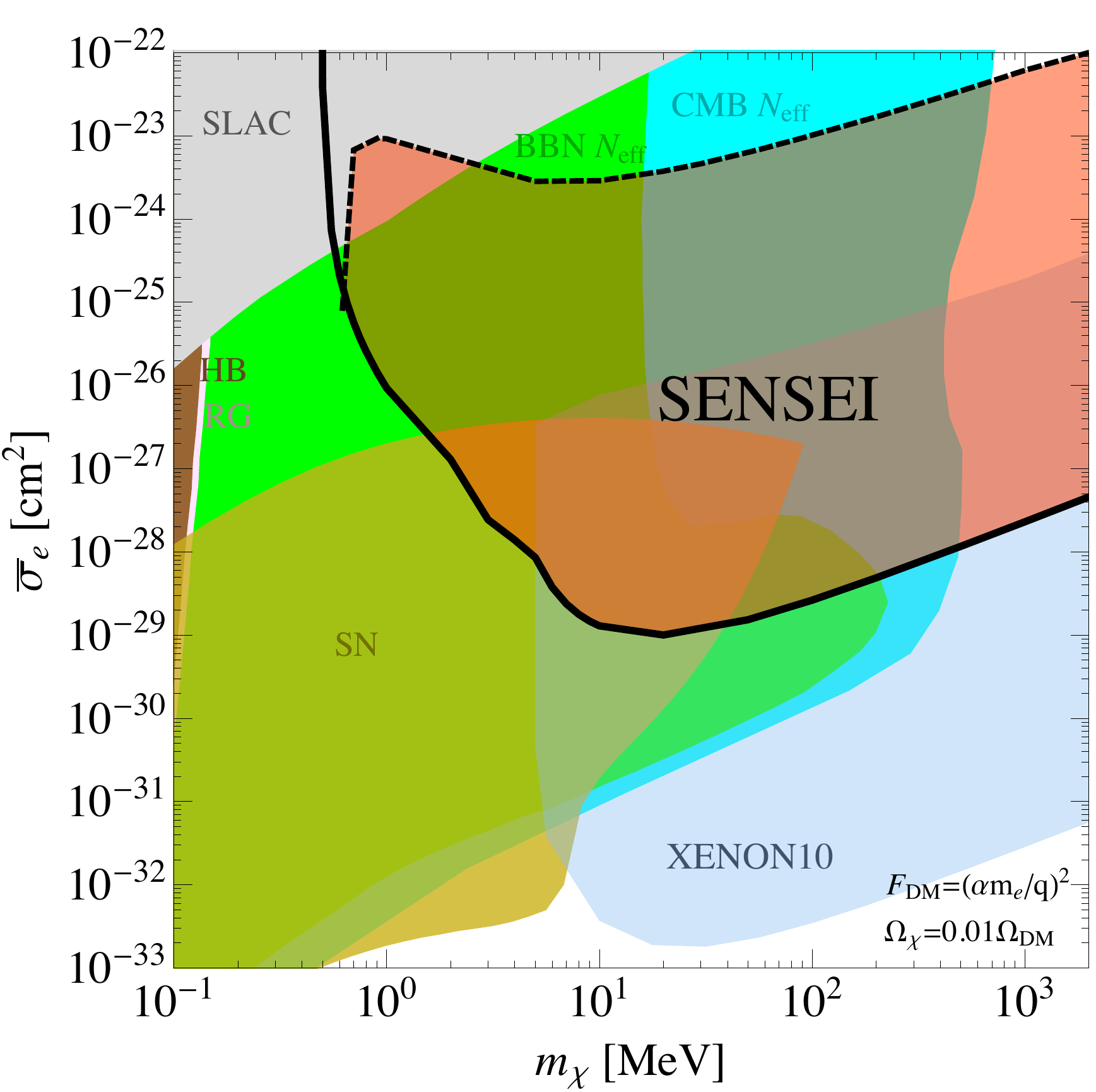}
\caption{
The 90\% C.L.~constraint on $\overline{\sigma}_e$ versus $m_\chi$ for $F_{\rm DM}(q^2) = (\alpha m_e/q)^2$ 
from a surface run at FNAL using the SENSEI prototype detector (orange region, bounded below by a solid line and above by 
a dashed line that is the same as the $|g_p|=|g_e|$ line in the right plot of Fig.~\ref{fig:DMescattering}). 
We assume that $\chi$ couples to an ultralight dark-photon mediator, and $\Omega_\chi=0.01\Omega_{\rm DM}$, 
which may explain the 21-cm signal observed by EDGES.  
Other 90\% C.L.~constraints are described in the text. 
The SENSEI surface run disfavors a small region of previously open parameter space for 
$\overline\sigma_e\gtrsim 10^{-25}~\textrm{cm}^2$ and $m_\chi$ greater than a few hundred MeV.  
}
\label{fig:DMsigmae}
\end{center}
\end{figure}

Our main results, for $\overline\sigma_e$ versus $m_\chi$ are shown in Fig.~\ref{fig:DMescattering} for the three form factors discussed above.  
Despite a small exposure time on the surface, the SENSEI commissioning run already probes novel parameter space 
for light DM (mass $\lesssim 4$~MeV) and for DM with large cross sections. 
This is the first time that a direct-detection constraint is derived for DM masses as low as $\sim$500~keV.  
In contrast, noble-liquid experiments (especially XENON10) probe lower cross sections for masses $\gtrsim 4$~MeV. 

In addition to having never probed DM masses below 4~MeV, the noble-liquid detectors that have previously constrained sub-GeV DM 
are operated underneath the Gran Sasso mountain. 
DM that interacts strongly with ordinary matter cannot reach these detectors due to scattering 
in the Earth. 
In contrast, much larger interaction strengths can be probed with the SENSEI surface run, as only the atmosphere (and a thin roof) 
can stop the DM. 

The terrestrial effects on MeV-to-GeV DM scattering off nuclei or electrons are model-dependent and 
have so far only been explored partially in the literature~\cite{Emken:2017erx,Emken:2017qmp,Kavanagh:2016pyr}; 
(see~\cite{Mack:2007xj,Kavanagh:2017cru,Davis:2017noy,Mahdawi:2017cxz,Mahdawi:2017utm,Hooper:2018bfw} 
for larger DM masses; see~\cite{An:2017ojc,Emken:2017hnp} for solar effects).  
However, to illustrate that SENSEI constrains novel parameter space at large cross sections, we include very preliminary results 
from~\cite{Emken:2018}.
Here, we estimate the terrestrial effects at the order-of-magnitude level. 
A dark-photon mediator or electric-dipole-moment allows DM to scatter off nuclei and electrons in the atmosphere or Earth 
(we include elastic scatters only, ignoring inelastic scatters off electrons). 
In the darker shaded regions in Fig.~\ref{fig:DMescattering} (labelled ``$|g_p|=|g_e|$'') the respective 
detectors have no sensitivity. 
If the mediator only couples to electrons (and not to nuclei), a very naive rescaling of the preliminary 
results in~\cite{Emken:2018} leads to the excluded regions labelled ``$g_p=0$'' (lighter shaded regions).  
If the mediator only couples to electrons (and not to nuclei), a naive estimate 
leads to the excluded regions labelled ``$g_p=0$'' (lighter shaded regions).  
We see that the SENSEI prototype constraints are largely complementary to existing noble-liquid detector constraints. 

\begin{figure}[t]
\begin{center}
\includegraphics[width=0.42\textwidth]{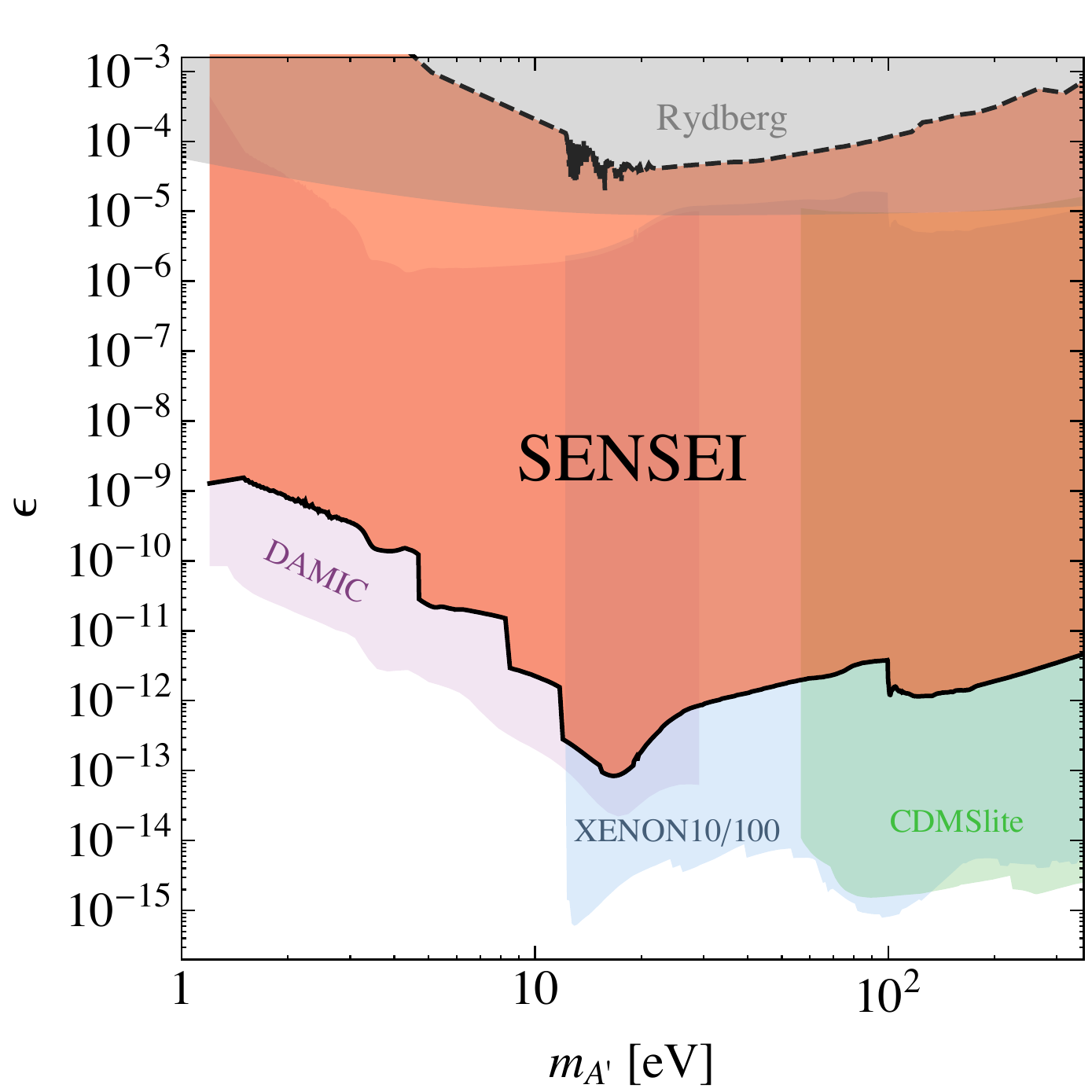}
\caption{90\% C.L.~constraints on dark-photon DM absorption by electrons, in the kinetic mixing ($\epsilon$) versus dark-photon mass 
($m_{A'}$) parameter space from a surface run at FNAL using the SENSEI prototype detector (orange). 
90\% C.L~constraints from XENON10, XENON100, CDMSlite, DAMIC, and a measurement of the Rydberg constant are shown. 
The direct-detection constraints are absent at large $\epsilon$, where the $A'$ is absorbed in the Earth's crust or atmosphere before 
reaching the detector.  
The SENSEI constraint closes a gap in laboratory probes of the high-$\epsilon$ region.   
We assume the $A'$ to be DM, irrespective of any (model-dependent) production process.  }
\label{fig:DMabsorption}
\end{center}
\end{figure}

We give one example of a concrete model that can give rise to large cross sections in Fig.~\ref{fig:DMsigmae}.  
We assume that a subdominant DM component, $\chi$, 
interacts with an ultralight dark photon ($m_{A'}\ll \textrm{keV}$), with $\Omega_\chi = 0.01 \Omega_{\rm DM}$.  
This model is motivated by the EDGES measurement of the 21-cm spectrum at $z\simeq17$, which revealed 
an anomalously large absorption signal~\cite{Bowman:2018yin}~\cite{Barkana:2018qrx} 
(see also~\cite{Tashiro:2014tsa,Munoz:2015bca,Barkana:2018lgd,Munoz:2018pzp,Berlin:2018sjs,
Fraser:2018acy,DAmico:2018sxd,Kang:2018qhi,Yang:2018gjd,Pospelov:2018kdh,Safarzadeh:2018hhg,Clark:2018ghm,Cheung:2018vww,Slatyer:2018aqg,Liu:2018uzy,Falkowski:2018qdj}). 
The SENSEI constraint (orange) is bounded by the solid (dashed) line for small (large~\cite{Emken:2018}) 
$\overline\sigma_e$.   
It disfavors novel parameter space for DM masses above a few hundred MeV for $\overline\sigma_e\gtrsim 10^{-25}~\textrm{cm}^2$. 
Other constraints arise from 
the SLAC millicharge experiment~\cite{Prinz:1998ua}, red-giant (RG) 
and horizontal-branch (HB) stars~\cite{Vogel:2013raa}, 
the BBN and CMB measurements of the number of relativistic degrees of freedom ($N_{\rm eff}$)~\cite{Vogel:2013raa}, 
and Supernova 1987A (SN)~\cite{Chang:2018rso}.  
A more careful analysis of other possible constraints in this region is, however, warranted, including 
a re-analysis of low-threshold DM-nuclear recoil data~\cite{Angloher:2017sxg,Petricca:2017zdp}, 
and an analysis of whether such a DM candidate 
would be evacuated from the Galactic disk by Galactic magnetic fields and supernova shock waves 
(as may be the case if the DM has a millicharge~\cite{McDermott:2010pa}).  

We next show the SENSEI prototype constraints on $\epsilon$ versus $m_{A'}$ for dark-photon DM ($A'$), which can be absorbed 
by an electron, in Fig.~\ref{fig:DMabsorption}.  
At small $\epsilon$, our new constraint is weaker than other constraints due to its small exposure, however, for large $\epsilon$ new grounds are explored.  We estimate the maximal coupling, $\epsilon_{\rm max}=1/\sqrt{\rho L\sigma_{\rm abs}}$ above which the $A'$ is absorbed 
by molecules in the atmosphere or by atomic electrons in the Earth's crust and sensitivity is lost. 
Here $\sigma_{\rm abs}$ is the measured photoabsorption cross-sections per molecule, $N$ is the average 
density, and $L$ is the depth.  
For simplicity, we take the Earth's crust to consist 
of silicon with $\rho= 2.7$~g/cm$^3$, with $L=$ 0.7~km (CDMSlite), 1.4~km (XENON10/100), or 2~km (DAMIC).  
We take the atmosphere to consist of O$_2$ and N$_2$~\cite{FENNELLY1992321} 
(with $\rho=1.2\times10^{-4}$ g/cm$^3$ and $L=$ 86~km).  These dominate the absorption for  
$m_{A'} \gtrsim 5$~eV,  the bond-dissociation energy of O$_2$ (no data on $\sigma_{\rm abs}$ is available below $m_{A'}\simeq 10$~eV, 
so we extrapolate the available data down to 5~eV).  For $m_{A'}\lesssim 5$~eV, ozone dominates the 
absorption, but its abundance is very small and does not affect the region shown in Fig.~\ref{fig:DMabsorption}.  
We see that SENSEI closes a gap in laboratory probes of the large-$\epsilon$ region: the gap was bounded above 
by a measurement of the Rydberg constant 
(we show a $2\sigma$ constraint adapted from~\cite{1999TJPh...23..943P,Karshenboim:2010ck,Karshenboim:2010cg}) 
and below by our analysis of terrestrial absorption effects for XENON10, XENON100, CDMSlite, and 
DAMIC~\cite{An:2014twa,Bloch:2016sjj,Hochberg:2016sqx,Aguilar-Arevalo:2016zop,Agnese:2015nto} (stellar constraints~\cite{An:2013yfc,Redondo:2013lna} already disfavor this region).

\section{Summary and Outlook} 

We present results from a low-exposure commissioning run of a SENSEI prototype detector.  We demonstrate the first use 
of Skipper-CCD technology for a DM search, the first direct-detection constraints for DM masses $0.5-4$~MeV, 
and the first direct-detection constraints on strongly interacting DM for masses between $0.5$~MeV and a few hundred MeV. 
Over the next few years, the SENSEI Collaboration aims to construct a detector consisting of $\sim$100~grams of Skipper CCDs 
that are fabricated in a dedicated production run using high-resistivity silicon.  These detectors are expected to have a dark current 
several orders of magnitude lower than the prototype detector and an improved single-sample noise.  
We expect to collect an exposure that is almost 2~million times larger than the exposure of the 
surface run and with far fewer background events, allowing us to explore vast new regions of DM parameter space.

\section{Acknowledgments}

We are grateful to Timon Emken, Chris Kouvaris, and Mukul Sholapurkar for extensive discussions and for letting us include 
preliminary results from the upcoming paper~\cite{Emken:2018}.  
We also thank Alex Drlica-Wagner and Erez Etzion for carefully reading a draft of the paper and for numerous useful comments.  
We are grateful for the support of the Heising-Simons Foundation under Grant No.~79921.
RE also acknowledges support from DoE Grant DE-SC0017938.  
This work was supported by Fermilab under DOE Contract No.\ DE-AC02-07CH11359. 
The work of TV is supported by the I-CORE Program of the Planning Budgeting Committee and the Israel Science Foundation (grant No.1937/12), by the European Research Council (ERC) under the EU Horizon 2020 Programme (ERC- CoG-2015 -Proposal n.~682676 LDMThExp), by the Israel Science Foundation-NSFC (grant No. 2522/17)  and by the German-Israeli Foundation (grant No. I-1283- 303.7/2014). TV is further supported by a grant from The Ambrose Monell Foundation, given by the Institute for Advanced Study.
RE and TV are also funded by the Binational Science Foundation (grant No. 2016153). 

\vskip10mm
{\bf Note added:}  
About one month after this paper appeared on the arxiv and while the paper was under journal review, the preprint~\cite{Agnese:2018col} 
appeared, showing constraints on DM-electron scattering that were obtained with a different technology and with different background challenges.  

 \bibliographystyle{apsrev4-1}
\bibliography{SENSEI.ScienceRun.2017.arxiv.v2.bbl}

\begin{thebibliography}{66}%
\makeatletter
\providecommand \@ifxundefined [1]{%
 \@ifx{#1\undefined}
}%
\providecommand \@ifnum [1]{%
 \ifnum #1\expandafter \@firstoftwo
 \else \expandafter \@secondoftwo
 \fi
}%
\providecommand \@ifx [1]{%
 \ifx #1\expandafter \@firstoftwo
 \else \expandafter \@secondoftwo
 \fi
}%
\providecommand \natexlab [1]{#1}%
\providecommand \enquote  [1]{``#1''}%
\providecommand \bibnamefont  [1]{#1}%
\providecommand \bibfnamefont [1]{#1}%
\providecommand \citenamefont [1]{#1}%
\providecommand \href@noop [0]{\@secondoftwo}%
\providecommand \href [0]{\begingroup \@sanitize@url \@href}%
\providecommand \@href[1]{\@@startlink{#1}\@@href}%
\providecommand \@@href[1]{\endgroup#1\@@endlink}%
\providecommand \@sanitize@url [0]{\catcode `\\12\catcode `\$12\catcode
  `\&12\catcode `\#12\catcode `\^12\catcode `\_12\catcode `\%12\relax}%
\providecommand \@@startlink[1]{}%
\providecommand \@@endlink[0]{}%
\providecommand \url  [0]{\begingroup\@sanitize@url \@url }%
\providecommand \@url [1]{\endgroup\@href {#1}{\urlprefix }}%
\providecommand \urlprefix  [0]{URL }%
\providecommand \Eprint [0]{\href }%
\providecommand \doibase [0]{http://dx.doi.org/}%
\providecommand \selectlanguage [0]{\@gobble}%
\providecommand \bibinfo  [0]{\@secondoftwo}%
\providecommand \bibfield  [0]{\@secondoftwo}%
\providecommand \translation [1]{[#1]}%
\providecommand \BibitemOpen [0]{}%
\providecommand \bibitemStop [0]{}%
\providecommand \bibitemNoStop [0]{.\EOS\space}%
\providecommand \EOS [0]{\spacefactor3000\relax}%
\providecommand \BibitemShut  [1]{\csname bibitem#1\endcsname}%
\let\auto@bib@innerbib\@empty
\bibitem [{\citenamefont {Battaglieri}\ \emph {et~al.}(2017)\citenamefont
  {Battaglieri} \emph {et~al.}}]{Battaglieri:2017aum}%
  \BibitemOpen
  \bibfield  {author} {\bibinfo {author} {\bibfnamefont {M.}~\bibnamefont
  {Battaglieri}} \emph {et~al.},\ }\href@noop {} {\  (\bibinfo {year}
  {2017})},\ \Eprint {http://arxiv.org/abs/1707.04591} {arXiv:1707.04591
  [hep-ph]} \BibitemShut {NoStop}%
\bibitem [{\citenamefont {Angloher}\ \emph {et~al.}(2017)\citenamefont
  {Angloher} \emph {et~al.}}]{Angloher:2017sxg}%
  \BibitemOpen
  \bibfield  {author} {\bibinfo {author} {\bibfnamefont {G.}~\bibnamefont
  {Angloher}} \emph {et~al.} (\bibinfo {collaboration} {CRESST}),\ }\href
  {\doibase 10.1140/epjc/s10052-017-5223-9} {\bibfield  {journal} {\bibinfo
  {journal} {Eur. Phys. J.}\ }\textbf {\bibinfo {volume} {C77}},\ \bibinfo
  {pages} {637} (\bibinfo {year} {2017})},\ \Eprint
  {http://arxiv.org/abs/1707.06749} {arXiv:1707.06749 [astro-ph.CO]}
  \BibitemShut {NoStop}%
\bibitem [{\citenamefont {Petricca}\ \emph {et~al.}(2017)\citenamefont
  {Petricca} \emph {et~al.}}]{Petricca:2017zdp}%
  \BibitemOpen
  \bibfield  {author} {\bibinfo {author} {\bibfnamefont {F.}~\bibnamefont
  {Petricca}} \emph {et~al.} (\bibinfo {collaboration} {CRESST}),\ }in\ \href
  {http://inspirehep.net/record/1637341/files/arXiv:1711.07692.pdf} {\emph
  {\bibinfo {booktitle} {{15th International Conference on Topics in
  Astroparticle and Underground Physics (TAUP 2017) Sudbury, Ontario, Canada,
  July 24-28, 2017}}}}\ (\bibinfo {year} {2017})\ \Eprint
  {http://arxiv.org/abs/1711.07692} {arXiv:1711.07692 [astro-ph.CO]}
  \BibitemShut {NoStop}%
\bibitem [{\citenamefont {Agnese}\ \emph {et~al.}(2016)\citenamefont {Agnese}
  \emph {et~al.}}]{Agnese:2015nto}%
  \BibitemOpen
  \bibfield  {author} {\bibinfo {author} {\bibfnamefont {R.}~\bibnamefont
  {Agnese}} \emph {et~al.} (\bibinfo {collaboration} {SuperCDMS}),\ }\href
  {\doibase 10.1103/PhysRevLett.116.071301} {\bibfield  {journal} {\bibinfo
  {journal} {Phys. Rev. Lett.}\ }\textbf {\bibinfo {volume} {116}},\ \bibinfo
  {pages} {071301} (\bibinfo {year} {2016})},\ \Eprint
  {http://arxiv.org/abs/1509.02448} {arXiv:1509.02448 [astro-ph.CO]}
  \BibitemShut {NoStop}%
\bibitem [{\citenamefont {Essig}\ \emph
  {et~al.}(2012{\natexlab{a}})\citenamefont {Essig}, \citenamefont {Mardon},\
  and\ \citenamefont {Volansky}}]{Essig:2011nj}%
  \BibitemOpen
  \bibfield  {author} {\bibinfo {author} {\bibfnamefont {R.}~\bibnamefont
  {Essig}}, \bibinfo {author} {\bibfnamefont {J.}~\bibnamefont {Mardon}}, \
  and\ \bibinfo {author} {\bibfnamefont {T.}~\bibnamefont {Volansky}},\ }\href
  {\doibase 10.1103/PhysRevD.85.076007} {\bibfield  {journal} {\bibinfo
  {journal} {Phys. Rev.}\ }\textbf {\bibinfo {volume} {D85}},\ \bibinfo {pages}
  {076007} (\bibinfo {year} {2012}{\natexlab{a}})},\ \Eprint
  {http://arxiv.org/abs/1108.5383} {arXiv:1108.5383 [hep-ph]} \BibitemShut
  {NoStop}%
\bibitem [{\citenamefont {Essig}\ \emph
  {et~al.}(2012{\natexlab{b}})\citenamefont {Essig}, \citenamefont
  {Manalaysay}, \citenamefont {Mardon}, \citenamefont {Sorensen},\ and\
  \citenamefont {Volansky}}]{Essig:2012yx}%
  \BibitemOpen
  \bibfield  {author} {\bibinfo {author} {\bibfnamefont {R.}~\bibnamefont
  {Essig}}, \bibinfo {author} {\bibfnamefont {A.}~\bibnamefont {Manalaysay}},
  \bibinfo {author} {\bibfnamefont {J.}~\bibnamefont {Mardon}}, \bibinfo
  {author} {\bibfnamefont {P.}~\bibnamefont {Sorensen}}, \ and\ \bibinfo
  {author} {\bibfnamefont {T.}~\bibnamefont {Volansky}},\ }\href {\doibase
  10.1103/PhysRevLett.109.021301} {\bibfield  {journal} {\bibinfo  {journal}
  {Phys. Rev. Lett.}\ }\textbf {\bibinfo {volume} {109}},\ \bibinfo {pages}
  {021301} (\bibinfo {year} {2012}{\natexlab{b}})},\ \Eprint
  {http://arxiv.org/abs/1206.2644} {arXiv:1206.2644 [astro-ph.CO]} \BibitemShut
  {NoStop}%
\bibitem [{\citenamefont {Essig}\ \emph {et~al.}(2017)\citenamefont {Essig},
  \citenamefont {Volansky},\ and\ \citenamefont {Yu}}]{Essig:2017kqs}%
  \BibitemOpen
  \bibfield  {author} {\bibinfo {author} {\bibfnamefont {R.}~\bibnamefont
  {Essig}}, \bibinfo {author} {\bibfnamefont {T.}~\bibnamefont {Volansky}}, \
  and\ \bibinfo {author} {\bibfnamefont {T.-T.}\ \bibnamefont {Yu}},\ }\href
  {\doibase 10.1103/PhysRevD.96.043017} {\bibfield  {journal} {\bibinfo
  {journal} {Phys. Rev.}\ }\textbf {\bibinfo {volume} {D96}},\ \bibinfo {pages}
  {043017} (\bibinfo {year} {2017})},\ \Eprint
  {http://arxiv.org/abs/1703.00910} {arXiv:1703.00910 [hep-ph]} \BibitemShut
  {NoStop}%
\bibitem [{\citenamefont {Angle}\ \emph {et~al.}(2011)\citenamefont {Angle}
  \emph {et~al.}}]{Angle:2011th}%
  \BibitemOpen
  \bibfield  {author} {\bibinfo {author} {\bibfnamefont {J.}~\bibnamefont
  {Angle}} \emph {et~al.} (\bibinfo {collaboration} {XENON10}),\ }\href
  {\doibase 10.1103/PhysRevLett.110.249901, 10.1103/PhysRevLett.107.051301}
  {\bibfield  {journal} {\bibinfo  {journal} {Phys. Rev. Lett.}\ }\textbf
  {\bibinfo {volume} {107}},\ \bibinfo {pages} {051301} (\bibinfo {year}
  {2011})},\ \bibinfo {note} {[Erratum: Phys. Rev. Lett.110,249901(2013)]},\
  \Eprint {http://arxiv.org/abs/1104.3088} {arXiv:1104.3088 [astro-ph.CO]}
  \BibitemShut {NoStop}%
\bibitem [{\citenamefont {Aprile}\ \emph {et~al.}(2016)\citenamefont {Aprile}
  \emph {et~al.}}]{Aprile:2016wwo}%
  \BibitemOpen
  \bibfield  {author} {\bibinfo {author} {\bibfnamefont {E.}~\bibnamefont
  {Aprile}} \emph {et~al.} (\bibinfo {collaboration} {XENON100}),\ }\href@noop
  {} {\  (\bibinfo {year} {2016})},\ \Eprint {http://arxiv.org/abs/1605.06262}
  {arXiv:1605.06262 [astro-ph.CO]} \BibitemShut {NoStop}%
\bibitem [{\citenamefont {Agnes}\ \emph {et~al.}(2018)\citenamefont {Agnes}
  \emph {et~al.}}]{Agnes:2018oej}%
  \BibitemOpen
  \bibfield  {author} {\bibinfo {author} {\bibfnamefont {P.}~\bibnamefont
  {Agnes}} \emph {et~al.} (\bibinfo {collaboration} {DarkSide}),\ }\href@noop
  {} {\  (\bibinfo {year} {2018})},\ \Eprint {http://arxiv.org/abs/1802.06998}
  {arXiv:1802.06998 [astro-ph.CO]} \BibitemShut {NoStop}%
\bibitem [{\citenamefont {Tiffenberg}\ \emph {et~al.}(2017)\citenamefont
  {Tiffenberg}, \citenamefont {Sofo-Haro}, \citenamefont {Drlica-Wagner},
  \citenamefont {Essig}, \citenamefont {Guardincerri}, \citenamefont {Holland},
  \citenamefont {Volansky},\ and\ \citenamefont {Yu}}]{Tiffenberg:2017aac}%
  \BibitemOpen
  \bibfield  {author} {\bibinfo {author} {\bibfnamefont {J.}~\bibnamefont
  {Tiffenberg}}, \bibinfo {author} {\bibfnamefont {M.}~\bibnamefont
  {Sofo-Haro}}, \bibinfo {author} {\bibfnamefont {A.}~\bibnamefont
  {Drlica-Wagner}}, \bibinfo {author} {\bibfnamefont {R.}~\bibnamefont
  {Essig}}, \bibinfo {author} {\bibfnamefont {Y.}~\bibnamefont {Guardincerri}},
  \bibinfo {author} {\bibfnamefont {S.}~\bibnamefont {Holland}}, \bibinfo
  {author} {\bibfnamefont {T.}~\bibnamefont {Volansky}}, \ and\ \bibinfo
  {author} {\bibfnamefont {T.-T.}\ \bibnamefont {Yu}},\ }\href {\doibase
  10.1103/PhysRevLett.119.131802} {\bibfield  {journal} {\bibinfo  {journal}
  {Phys. Rev. Lett.}\ }\textbf {\bibinfo {volume} {119}},\ \bibinfo {pages}
  {131802} (\bibinfo {year} {2017})},\ \Eprint
  {http://arxiv.org/abs/1706.00028} {arXiv:1706.00028 [physics.ins-det]}
  \BibitemShut {NoStop}%
\bibitem [{\citenamefont {Romani}\ \emph {et~al.}(2017)\citenamefont {Romani}
  \emph {et~al.}}]{Romani:2017iwi}%
  \BibitemOpen
  \bibfield  {author} {\bibinfo {author} {\bibfnamefont {R.~K.}\ \bibnamefont
  {Romani}} \emph {et~al.},\ }\href@noop {} {\  (\bibinfo {year} {2017})},\
  \Eprint {http://arxiv.org/abs/1710.09335} {arXiv:1710.09335
  [physics.ins-det]} \BibitemShut {NoStop}%
\bibitem [{\citenamefont {Essig}\ \emph {et~al.}(2016)\citenamefont {Essig},
  \citenamefont {Fernandez-Serra}, \citenamefont {Mardon}, \citenamefont
  {Soto}, \citenamefont {Volansky},\ and\ \citenamefont {Yu}}]{Essig:2015cda}%
  \BibitemOpen
  \bibfield  {author} {\bibinfo {author} {\bibfnamefont {R.}~\bibnamefont
  {Essig}}, \bibinfo {author} {\bibfnamefont {M.}~\bibnamefont
  {Fernandez-Serra}}, \bibinfo {author} {\bibfnamefont {J.}~\bibnamefont
  {Mardon}}, \bibinfo {author} {\bibfnamefont {A.}~\bibnamefont {Soto}},
  \bibinfo {author} {\bibfnamefont {T.}~\bibnamefont {Volansky}}, \ and\
  \bibinfo {author} {\bibfnamefont {T.-T.}\ \bibnamefont {Yu}},\ }\href
  {\doibase 10.1007/JHEP05(2016)046} {\bibfield  {journal} {\bibinfo  {journal}
  {JHEP}\ }\textbf {\bibinfo {volume} {05}},\ \bibinfo {pages} {046} (\bibinfo
  {year} {2016})},\ \Eprint {http://arxiv.org/abs/1509.01598} {arXiv:1509.01598
  [hep-ph]} \BibitemShut {NoStop}%
\bibitem [{\citenamefont {Lee}\ \emph {et~al.}(2015)\citenamefont {Lee},
  \citenamefont {Lisanti}, \citenamefont {Mishra-Sharma},\ and\ \citenamefont
  {Safdi}}]{Lee:2015qva}%
  \BibitemOpen
  \bibfield  {author} {\bibinfo {author} {\bibfnamefont {S.~K.}\ \bibnamefont
  {Lee}}, \bibinfo {author} {\bibfnamefont {M.}~\bibnamefont {Lisanti}},
  \bibinfo {author} {\bibfnamefont {S.}~\bibnamefont {Mishra-Sharma}}, \ and\
  \bibinfo {author} {\bibfnamefont {B.~R.}\ \bibnamefont {Safdi}},\ }\href
  {\doibase 10.1103/PhysRevD.92.083517} {\bibfield  {journal} {\bibinfo
  {journal} {Phys. Rev.}\ }\textbf {\bibinfo {volume} {D92}},\ \bibinfo {pages}
  {083517} (\bibinfo {year} {2015})},\ \Eprint
  {http://arxiv.org/abs/1508.07361} {arXiv:1508.07361 [hep-ph]} \BibitemShut
  {NoStop}%
\bibitem [{\citenamefont {Graham}\ \emph {et~al.}(2012)\citenamefont {Graham},
  \citenamefont {Kaplan}, \citenamefont {Rajendran},\ and\ \citenamefont
  {Walters}}]{Graham:2012su}%
  \BibitemOpen
  \bibfield  {author} {\bibinfo {author} {\bibfnamefont {P.~W.}\ \bibnamefont
  {Graham}}, \bibinfo {author} {\bibfnamefont {D.~E.}\ \bibnamefont {Kaplan}},
  \bibinfo {author} {\bibfnamefont {S.}~\bibnamefont {Rajendran}}, \ and\
  \bibinfo {author} {\bibfnamefont {M.~T.}\ \bibnamefont {Walters}},\ }\href
  {\doibase 10.1016/j.dark.2012.09.001} {\bibfield  {journal} {\bibinfo
  {journal} {Phys.Dark Univ.}\ }\textbf {\bibinfo {volume} {1}},\ \bibinfo
  {pages} {32} (\bibinfo {year} {2012})},\ \Eprint
  {http://arxiv.org/abs/1203.2531} {arXiv:1203.2531 [hep-ph]} \BibitemShut
  {NoStop}%
\bibitem [{Note1()}]{Note1}%
  \BibitemOpen
  \bibinfo {note} {The CCD temperature is measured with an RTD placed some
  distance away, resulting in an uncertainty of a few degrees.}\BibitemShut
  {Stop}%
\bibitem [{\citenamefont {Emken}\ \emph {et~al.}()\citenamefont {Emken},
  \citenamefont {Essig}, \citenamefont {Kouvaris},\ and\ \citenamefont
  {Sholapurkar}}]{Emken:2018}%
  \BibitemOpen
  \bibfield  {author} {\bibinfo {author} {\bibfnamefont {T.}~\bibnamefont
  {Emken}}, \bibinfo {author} {\bibfnamefont {R.}~\bibnamefont {Essig}},
  \bibinfo {author} {\bibfnamefont {C.}~\bibnamefont {Kouvaris}}, \ and\
  \bibinfo {author} {\bibfnamefont {M.}~\bibnamefont {Sholapurkar}},\
  }\href@noop {} {\ }\Eprint {http://arxiv.org/abs/to appear} {to appear}
  \BibitemShut {NoStop}%
\bibitem [{\citenamefont {Aguilar-Arevalo}\ \emph
  {et~al.}(2016{\natexlab{a}})\citenamefont {Aguilar-Arevalo} \emph
  {et~al.}}]{Aguilar-Arevalo:2016khx}%
  \BibitemOpen
  \bibfield  {author} {\bibinfo {author} {\bibfnamefont {A.}~\bibnamefont
  {Aguilar-Arevalo}} \emph {et~al.} (\bibinfo {collaboration} {CONNIE}),\
  }\bibfield  {booktitle} {\emph {\bibinfo {booktitle} {{Proceedings, 15th
  Mexican Workshop on Particles and Fields (MWPF 2015): Mazatl\'an, M\'exico,
  November 2-6, 2015}}},\ }\href {\doibase 10.1088/1742-6596/761/1/012057}
  {\bibfield  {journal} {\bibinfo  {journal} {J. Phys. Conf. Ser.}\ }\textbf
  {\bibinfo {volume} {761}},\ \bibinfo {pages} {012057} (\bibinfo {year}
  {2016}{\natexlab{a}})},\ \Eprint {http://arxiv.org/abs/1608.01565}
  {arXiv:1608.01565 [physics.ins-det]} \BibitemShut {NoStop}%
\bibitem [{\citenamefont {Aguilar-Arevalo}\ \emph
  {et~al.}(2016{\natexlab{b}})\citenamefont {Aguilar-Arevalo} \emph
  {et~al.}}]{Aguilar-Arevalo:2016ndq}%
  \BibitemOpen
  \bibfield  {author} {\bibinfo {author} {\bibfnamefont {A.}~\bibnamefont
  {Aguilar-Arevalo}} \emph {et~al.} (\bibinfo {collaboration} {DAMIC}),\ }\href
  {\doibase 10.1103/PhysRevD.94.082006} {\bibfield  {journal} {\bibinfo
  {journal} {Phys. Rev.}\ }\textbf {\bibinfo {volume} {D94}},\ \bibinfo {pages}
  {082006} (\bibinfo {year} {2016}{\natexlab{b}})},\ \Eprint
  {http://arxiv.org/abs/1607.07410} {arXiv:1607.07410 [astro-ph.CO]}
  \BibitemShut {NoStop}%
\bibitem [{\citenamefont {Holland}\ \emph {et~al.}(2003)\citenamefont
  {Holland}, \citenamefont {Groom}, \citenamefont {Palaio}, \citenamefont
  {Stover},\ and\ \citenamefont {Wei}}]{Holland:2003}%
  \BibitemOpen
  \bibfield  {author} {\bibinfo {author} {\bibfnamefont {S.~E.}\ \bibnamefont
  {Holland}}, \bibinfo {author} {\bibfnamefont {D.~E.}\ \bibnamefont {Groom}},
  \bibinfo {author} {\bibfnamefont {N.~P.}\ \bibnamefont {Palaio}}, \bibinfo
  {author} {\bibfnamefont {R.~J.}\ \bibnamefont {Stover}}, \ and\ \bibinfo
  {author} {\bibfnamefont {M.}~\bibnamefont {Wei}},\ }\href {\doibase
  10.1109/TED.2002.806476} {\bibfield  {journal} {\bibinfo  {journal} {IEEE
  Transactions on Electron Devices}\ }\textbf {\bibinfo {volume} {50}},\
  \bibinfo {pages} {225} (\bibinfo {year} {2003})}\BibitemShut {NoStop}%
\bibitem [{\citenamefont {An}\ \emph {et~al.}(2015)\citenamefont {An},
  \citenamefont {Pospelov}, \citenamefont {Pradler},\ and\ \citenamefont
  {Ritz}}]{An:2014twa}%
  \BibitemOpen
  \bibfield  {author} {\bibinfo {author} {\bibfnamefont {H.}~\bibnamefont
  {An}}, \bibinfo {author} {\bibfnamefont {M.}~\bibnamefont {Pospelov}},
  \bibinfo {author} {\bibfnamefont {J.}~\bibnamefont {Pradler}}, \ and\
  \bibinfo {author} {\bibfnamefont {A.}~\bibnamefont {Ritz}},\ }\href {\doibase
  10.1016/j.physletb.2015.06.018} {\bibfield  {journal} {\bibinfo  {journal}
  {Phys. Lett.}\ }\textbf {\bibinfo {volume} {B747}},\ \bibinfo {pages} {331}
  (\bibinfo {year} {2015})},\ \Eprint {http://arxiv.org/abs/1412.8378}
  {arXiv:1412.8378 [hep-ph]} \BibitemShut {NoStop}%
\bibitem [{\citenamefont {Bloch}\ \emph {et~al.}(2017)\citenamefont {Bloch},
  \citenamefont {Essig}, \citenamefont {Tobioka}, \citenamefont {Volansky},\
  and\ \citenamefont {Yu}}]{Bloch:2016sjj}%
  \BibitemOpen
  \bibfield  {author} {\bibinfo {author} {\bibfnamefont {I.~M.}\ \bibnamefont
  {Bloch}}, \bibinfo {author} {\bibfnamefont {R.}~\bibnamefont {Essig}},
  \bibinfo {author} {\bibfnamefont {K.}~\bibnamefont {Tobioka}}, \bibinfo
  {author} {\bibfnamefont {T.}~\bibnamefont {Volansky}}, \ and\ \bibinfo
  {author} {\bibfnamefont {T.-T.}\ \bibnamefont {Yu}},\ }\href {\doibase
  10.1007/JHEP06(2017)087} {\bibfield  {journal} {\bibinfo  {journal} {JHEP}\
  }\textbf {\bibinfo {volume} {06}},\ \bibinfo {pages} {087} (\bibinfo {year}
  {2017})},\ \Eprint {http://arxiv.org/abs/1608.02123} {arXiv:1608.02123
  [hep-ph]} \BibitemShut {NoStop}%
\bibitem [{\citenamefont {Hochberg}\ \emph {et~al.}(2017)\citenamefont
  {Hochberg}, \citenamefont {Lin},\ and\ \citenamefont
  {Zurek}}]{Hochberg:2016sqx}%
  \BibitemOpen
  \bibfield  {author} {\bibinfo {author} {\bibfnamefont {Y.}~\bibnamefont
  {Hochberg}}, \bibinfo {author} {\bibfnamefont {T.}~\bibnamefont {Lin}}, \
  and\ \bibinfo {author} {\bibfnamefont {K.~M.}\ \bibnamefont {Zurek}},\ }\href
  {\doibase 10.1103/PhysRevD.95.023013} {\bibfield  {journal} {\bibinfo
  {journal} {Phys. Rev.}\ }\textbf {\bibinfo {volume} {D95}},\ \bibinfo {pages}
  {023013} (\bibinfo {year} {2017})},\ \Eprint
  {http://arxiv.org/abs/1608.01994} {arXiv:1608.01994 [hep-ph]} \BibitemShut
  {NoStop}%
\bibitem [{\citenamefont {Hochberg}\ \emph {et~al.}(2016)\citenamefont
  {Hochberg}, \citenamefont {Lin},\ and\ \citenamefont
  {Zurek}}]{Hochberg:2016ajh}%
  \BibitemOpen
  \bibfield  {author} {\bibinfo {author} {\bibfnamefont {Y.}~\bibnamefont
  {Hochberg}}, \bibinfo {author} {\bibfnamefont {T.}~\bibnamefont {Lin}}, \
  and\ \bibinfo {author} {\bibfnamefont {K.~M.}\ \bibnamefont {Zurek}},\ }\href
  {\doibase 10.1103/PhysRevD.94.015019} {\bibfield  {journal} {\bibinfo
  {journal} {Phys. Rev.}\ }\textbf {\bibinfo {volume} {D94}},\ \bibinfo {pages}
  {015019} (\bibinfo {year} {2016})},\ \Eprint
  {http://arxiv.org/abs/1604.06800} {arXiv:1604.06800 [hep-ph]} \BibitemShut
  {NoStop}%
\bibitem [{\citenamefont {Salucci}\ \emph {et~al.}(2010)\citenamefont
  {Salucci}, \citenamefont {Nesti}, \citenamefont {Gentile},\ and\
  \citenamefont {Martins}}]{Salucci:2010qr}%
  \BibitemOpen
  \bibfield  {author} {\bibinfo {author} {\bibfnamefont {P.}~\bibnamefont
  {Salucci}}, \bibinfo {author} {\bibfnamefont {F.}~\bibnamefont {Nesti}},
  \bibinfo {author} {\bibfnamefont {G.}~\bibnamefont {Gentile}}, \ and\
  \bibinfo {author} {\bibfnamefont {C.~F.}\ \bibnamefont {Martins}},\ }\href
  {\doibase 10.1051/0004-6361/201014385} {\bibfield  {journal} {\bibinfo
  {journal} {Astron. Astrophys.}\ }\textbf {\bibinfo {volume} {523}},\ \bibinfo
  {pages} {A83} (\bibinfo {year} {2010})},\ \Eprint
  {http://arxiv.org/abs/1003.3101} {arXiv:1003.3101 [astro-ph.GA]} \BibitemShut
  {NoStop}%
\bibitem [{\citenamefont {Emken}\ \emph {et~al.}(2017)\citenamefont {Emken},
  \citenamefont {Kouvaris},\ and\ \citenamefont {Shoemaker}}]{Emken:2017erx}%
  \BibitemOpen
  \bibfield  {author} {\bibinfo {author} {\bibfnamefont {T.}~\bibnamefont
  {Emken}}, \bibinfo {author} {\bibfnamefont {C.}~\bibnamefont {Kouvaris}}, \
  and\ \bibinfo {author} {\bibfnamefont {I.~M.}\ \bibnamefont {Shoemaker}},\
  }\href {\doibase 10.1103/PhysRevD.96.015018} {\bibfield  {journal} {\bibinfo
  {journal} {Phys. Rev.}\ }\textbf {\bibinfo {volume} {D96}},\ \bibinfo {pages}
  {015018} (\bibinfo {year} {2017})},\ \Eprint
  {http://arxiv.org/abs/1702.07750} {arXiv:1702.07750 [hep-ph]} \BibitemShut
  {NoStop}%
\bibitem [{\citenamefont {Emken}\ and\ \citenamefont
  {Kouvaris}(2017)}]{Emken:2017qmp}%
  \BibitemOpen
  \bibfield  {author} {\bibinfo {author} {\bibfnamefont {T.}~\bibnamefont
  {Emken}}\ and\ \bibinfo {author} {\bibfnamefont {C.}~\bibnamefont
  {Kouvaris}},\ }\href {\doibase 10.1088/1475-7516/2017/10/031} {\bibfield
  {journal} {\bibinfo  {journal} {JCAP}\ }\textbf {\bibinfo {volume} {1710}},\
  \bibinfo {pages} {031} (\bibinfo {year} {2017})},\ \Eprint
  {http://arxiv.org/abs/1706.02249} {arXiv:1706.02249 [hep-ph]} \BibitemShut
  {NoStop}%
\bibitem [{\citenamefont {Kavanagh}\ \emph {et~al.}(2017)\citenamefont
  {Kavanagh}, \citenamefont {Catena},\ and\ \citenamefont
  {Kouvaris}}]{Kavanagh:2016pyr}%
  \BibitemOpen
  \bibfield  {author} {\bibinfo {author} {\bibfnamefont {B.~J.}\ \bibnamefont
  {Kavanagh}}, \bibinfo {author} {\bibfnamefont {R.}~\bibnamefont {Catena}}, \
  and\ \bibinfo {author} {\bibfnamefont {C.}~\bibnamefont {Kouvaris}},\ }\href
  {\doibase 10.1088/1475-7516/2017/01/012} {\bibfield  {journal} {\bibinfo
  {journal} {JCAP}\ }\textbf {\bibinfo {volume} {1701}},\ \bibinfo {pages}
  {012} (\bibinfo {year} {2017})},\ \Eprint {http://arxiv.org/abs/1611.05453}
  {arXiv:1611.05453 [hep-ph]} \BibitemShut {NoStop}%
\bibitem [{\citenamefont {Mack}\ \emph {et~al.}(2007)\citenamefont {Mack},
  \citenamefont {Beacom},\ and\ \citenamefont {Bertone}}]{Mack:2007xj}%
  \BibitemOpen
  \bibfield  {author} {\bibinfo {author} {\bibfnamefont {G.~D.}\ \bibnamefont
  {Mack}}, \bibinfo {author} {\bibfnamefont {J.~F.}\ \bibnamefont {Beacom}}, \
  and\ \bibinfo {author} {\bibfnamefont {G.}~\bibnamefont {Bertone}},\ }\href
  {\doibase 10.1103/PhysRevD.76.043523} {\bibfield  {journal} {\bibinfo
  {journal} {Phys. Rev.}\ }\textbf {\bibinfo {volume} {D76}},\ \bibinfo {pages}
  {043523} (\bibinfo {year} {2007})},\ \Eprint {http://arxiv.org/abs/0705.4298}
  {arXiv:0705.4298 [astro-ph]} \BibitemShut {NoStop}%
\bibitem [{\citenamefont {Kavanagh}(2017)}]{Kavanagh:2017cru}%
  \BibitemOpen
  \bibfield  {author} {\bibinfo {author} {\bibfnamefont {B.~J.}\ \bibnamefont
  {Kavanagh}},\ }\href@noop {} {\  (\bibinfo {year} {2017})},\ \Eprint
  {http://arxiv.org/abs/1712.04901} {arXiv:1712.04901 [hep-ph]} \BibitemShut
  {NoStop}%
\bibitem [{\citenamefont {Davis}(2017)}]{Davis:2017noy}%
  \BibitemOpen
  \bibfield  {author} {\bibinfo {author} {\bibfnamefont {J.~H.}\ \bibnamefont
  {Davis}},\ }\href {\doibase 10.1103/PhysRevLett.119.211302} {\bibfield
  {journal} {\bibinfo  {journal} {Phys. Rev. Lett.}\ }\textbf {\bibinfo
  {volume} {119}},\ \bibinfo {pages} {211302} (\bibinfo {year} {2017})},\
  \Eprint {http://arxiv.org/abs/1708.01484} {arXiv:1708.01484 [hep-ph]}
  \BibitemShut {NoStop}%
\bibitem [{\citenamefont {Mahdawi}\ and\ \citenamefont
  {Farrar}(2017{\natexlab{a}})}]{Mahdawi:2017cxz}%
  \BibitemOpen
  \bibfield  {author} {\bibinfo {author} {\bibfnamefont {M.~S.}\ \bibnamefont
  {Mahdawi}}\ and\ \bibinfo {author} {\bibfnamefont {G.~R.}\ \bibnamefont
  {Farrar}},\ }\href {\doibase 10.1088/1475-7516/2017/12/004} {\bibfield
  {journal} {\bibinfo  {journal} {JCAP}\ }\textbf {\bibinfo {volume} {1712}},\
  \bibinfo {pages} {004} (\bibinfo {year} {2017}{\natexlab{a}})},\ \Eprint
  {http://arxiv.org/abs/1709.00430} {arXiv:1709.00430 [hep-ph]} \BibitemShut
  {NoStop}%
\bibitem [{\citenamefont {Mahdawi}\ and\ \citenamefont
  {Farrar}(2017{\natexlab{b}})}]{Mahdawi:2017utm}%
  \BibitemOpen
  \bibfield  {author} {\bibinfo {author} {\bibfnamefont {M.~S.}\ \bibnamefont
  {Mahdawi}}\ and\ \bibinfo {author} {\bibfnamefont {G.~R.}\ \bibnamefont
  {Farrar}},\ }\href@noop {} {\  (\bibinfo {year} {2017}{\natexlab{b}})},\
  \Eprint {http://arxiv.org/abs/1712.01170} {arXiv:1712.01170 [hep-ph]}
  \BibitemShut {NoStop}%
\bibitem [{\citenamefont {Hooper}\ and\ \citenamefont
  {McDermott}(2018)}]{Hooper:2018bfw}%
  \BibitemOpen
  \bibfield  {author} {\bibinfo {author} {\bibfnamefont {D.}~\bibnamefont
  {Hooper}}\ and\ \bibinfo {author} {\bibfnamefont {S.~D.}\ \bibnamefont
  {McDermott}},\ }\href@noop {} {\  (\bibinfo {year} {2018})},\ \Eprint
  {http://arxiv.org/abs/1802.03025} {arXiv:1802.03025 [hep-ph]} \BibitemShut
  {NoStop}%
\bibitem [{\citenamefont {An}\ \emph {et~al.}(2018)\citenamefont {An},
  \citenamefont {Pospelov}, \citenamefont {Pradler},\ and\ \citenamefont
  {Ritz}}]{An:2017ojc}%
  \BibitemOpen
  \bibfield  {author} {\bibinfo {author} {\bibfnamefont {H.}~\bibnamefont
  {An}}, \bibinfo {author} {\bibfnamefont {M.}~\bibnamefont {Pospelov}},
  \bibinfo {author} {\bibfnamefont {J.}~\bibnamefont {Pradler}}, \ and\
  \bibinfo {author} {\bibfnamefont {A.}~\bibnamefont {Ritz}},\ }\href {\doibase
  10.1103/PhysRevLett.120.141801} {\bibfield  {journal} {\bibinfo  {journal}
  {Phys. Rev. Lett.}\ }\textbf {\bibinfo {volume} {120}},\ \bibinfo {pages}
  {141801} (\bibinfo {year} {2018})},\ \Eprint
  {http://arxiv.org/abs/1708.03642} {arXiv:1708.03642 [hep-ph]} \BibitemShut
  {NoStop}%
\bibitem [{\citenamefont {Emken}\ \emph {et~al.}(2018)\citenamefont {Emken},
  \citenamefont {Kouvaris},\ and\ \citenamefont {Nielsen}}]{Emken:2017hnp}%
  \BibitemOpen
  \bibfield  {author} {\bibinfo {author} {\bibfnamefont {T.}~\bibnamefont
  {Emken}}, \bibinfo {author} {\bibfnamefont {C.}~\bibnamefont {Kouvaris}}, \
  and\ \bibinfo {author} {\bibfnamefont {N.~G.}\ \bibnamefont {Nielsen}},\
  }\href {\doibase 10.1103/PhysRevD.97.063007} {\bibfield  {journal} {\bibinfo
  {journal} {Phys. Rev.}\ }\textbf {\bibinfo {volume} {D97}},\ \bibinfo {pages}
  {063007} (\bibinfo {year} {2018})},\ \Eprint
  {http://arxiv.org/abs/1709.06573} {arXiv:1709.06573 [hep-ph]} \BibitemShut
  {NoStop}%
\bibitem [{\citenamefont {Bowman}\ \emph {et~al.}(2018)\citenamefont {Bowman},
  \citenamefont {Rogers}, \citenamefont {Monsalve}, \citenamefont {Mozdzen},\
  and\ \citenamefont {Mahesh}}]{Bowman:2018yin}%
  \BibitemOpen
  \bibfield  {author} {\bibinfo {author} {\bibfnamefont {J.~D.}\ \bibnamefont
  {Bowman}}, \bibinfo {author} {\bibfnamefont {A.~E.~E.}\ \bibnamefont
  {Rogers}}, \bibinfo {author} {\bibfnamefont {R.~A.}\ \bibnamefont
  {Monsalve}}, \bibinfo {author} {\bibfnamefont {T.~J.}\ \bibnamefont
  {Mozdzen}}, \ and\ \bibinfo {author} {\bibfnamefont {N.}~\bibnamefont
  {Mahesh}},\ }\href {\doibase 10.1038/nature25792} {\bibfield  {journal}
  {\bibinfo  {journal} {Nature}\ }\textbf {\bibinfo {volume} {555}},\ \bibinfo
  {pages} {67} (\bibinfo {year} {2018})}\BibitemShut {NoStop}%
\bibitem [{\citenamefont {Barkana}\ \emph {et~al.}(2018)\citenamefont
  {Barkana}, \citenamefont {Outmezguine}, \citenamefont {Redigolo},\ and\
  \citenamefont {Volansky}}]{Barkana:2018qrx}%
  \BibitemOpen
  \bibfield  {author} {\bibinfo {author} {\bibfnamefont {R.}~\bibnamefont
  {Barkana}}, \bibinfo {author} {\bibfnamefont {N.~J.}\ \bibnamefont
  {Outmezguine}}, \bibinfo {author} {\bibfnamefont {D.}~\bibnamefont
  {Redigolo}}, \ and\ \bibinfo {author} {\bibfnamefont {T.}~\bibnamefont
  {Volansky}},\ }\href@noop {} {\  (\bibinfo {year} {2018})},\ \Eprint
  {http://arxiv.org/abs/1803.03091} {arXiv:1803.03091 [hep-ph]} \BibitemShut
  {NoStop}%
\bibitem [{\citenamefont {Tashiro}\ \emph {et~al.}(2014)\citenamefont
  {Tashiro}, \citenamefont {Kadota},\ and\ \citenamefont
  {Silk}}]{Tashiro:2014tsa}%
  \BibitemOpen
  \bibfield  {author} {\bibinfo {author} {\bibfnamefont {H.}~\bibnamefont
  {Tashiro}}, \bibinfo {author} {\bibfnamefont {K.}~\bibnamefont {Kadota}}, \
  and\ \bibinfo {author} {\bibfnamefont {J.}~\bibnamefont {Silk}},\ }\href
  {\doibase 10.1103/PhysRevD.90.083522} {\bibfield  {journal} {\bibinfo
  {journal} {Phys. Rev.}\ }\textbf {\bibinfo {volume} {D90}},\ \bibinfo {pages}
  {083522} (\bibinfo {year} {2014})},\ \Eprint {http://arxiv.org/abs/1408.2571}
  {arXiv:1408.2571 [astro-ph.CO]} \BibitemShut {NoStop}%
\bibitem [{\citenamefont {Muñoz}\ \emph {et~al.}(2015)\citenamefont {Muñoz},
  \citenamefont {Kovetz},\ and\ \citenamefont {Ali-Haïmoud}}]{Munoz:2015bca}%
  \BibitemOpen
  \bibfield  {author} {\bibinfo {author} {\bibfnamefont {J.~B.}\ \bibnamefont
  {Muñoz}}, \bibinfo {author} {\bibfnamefont {E.~D.}\ \bibnamefont {Kovetz}},
  \ and\ \bibinfo {author} {\bibfnamefont {Y.}~\bibnamefont {Ali-Haïmoud}},\
  }\href {\doibase 10.1103/PhysRevD.92.083528} {\bibfield  {journal} {\bibinfo
  {journal} {Phys. Rev.}\ }\textbf {\bibinfo {volume} {D92}},\ \bibinfo {pages}
  {083528} (\bibinfo {year} {2015})},\ \Eprint
  {http://arxiv.org/abs/1509.00029} {arXiv:1509.00029 [astro-ph.CO]}
  \BibitemShut {NoStop}%
\bibitem [{\citenamefont {Barkana}(2018)}]{Barkana:2018lgd}%
  \BibitemOpen
  \bibfield  {author} {\bibinfo {author} {\bibfnamefont {R.}~\bibnamefont
  {Barkana}},\ }\href {\doibase 10.1038/nature25791} {\bibfield  {journal}
  {\bibinfo  {journal} {Nature}\ }\textbf {\bibinfo {volume} {555}},\ \bibinfo
  {pages} {71} (\bibinfo {year} {2018})},\ \Eprint
  {http://arxiv.org/abs/1803.06698} {arXiv:1803.06698 [astro-ph.CO]}
  \BibitemShut {NoStop}%
\bibitem [{\citenamefont {Muñoz}\ and\ \citenamefont
  {Loeb}(2018)}]{Munoz:2018pzp}%
  \BibitemOpen
  \bibfield  {author} {\bibinfo {author} {\bibfnamefont {J.~B.}\ \bibnamefont
  {Muñoz}}\ and\ \bibinfo {author} {\bibfnamefont {A.}~\bibnamefont {Loeb}},\
  }\href@noop {} {\  (\bibinfo {year} {2018})},\ \Eprint
  {http://arxiv.org/abs/1802.10094} {arXiv:1802.10094 [astro-ph.CO]}
  \BibitemShut {NoStop}%
\bibitem [{\citenamefont {Berlin}\ \emph {et~al.}(2018)\citenamefont {Berlin},
  \citenamefont {Hooper}, \citenamefont {Krnjaic},\ and\ \citenamefont
  {McDermott}}]{Berlin:2018sjs}%
  \BibitemOpen
  \bibfield  {author} {\bibinfo {author} {\bibfnamefont {A.}~\bibnamefont
  {Berlin}}, \bibinfo {author} {\bibfnamefont {D.}~\bibnamefont {Hooper}},
  \bibinfo {author} {\bibfnamefont {G.}~\bibnamefont {Krnjaic}}, \ and\
  \bibinfo {author} {\bibfnamefont {S.~D.}\ \bibnamefont {McDermott}},\
  }\href@noop {} {\  (\bibinfo {year} {2018})},\ \Eprint
  {http://arxiv.org/abs/1803.02804} {arXiv:1803.02804 [hep-ph]} \BibitemShut
  {NoStop}%
\bibitem [{\citenamefont {Fraser}\ \emph {et~al.}(2018)\citenamefont {Fraser}
  \emph {et~al.}}]{Fraser:2018acy}%
  \BibitemOpen
  \bibfield  {author} {\bibinfo {author} {\bibfnamefont {S.}~\bibnamefont
  {Fraser}} \emph {et~al.},\ }\href@noop {} {\  (\bibinfo {year} {2018})},\
  \Eprint {http://arxiv.org/abs/1803.03245} {arXiv:1803.03245 [hep-ph]}
  \BibitemShut {NoStop}%
\bibitem [{\citenamefont {D'Amico}\ \emph {et~al.}(2018)\citenamefont
  {D'Amico}, \citenamefont {Panci},\ and\ \citenamefont
  {Strumia}}]{DAmico:2018sxd}%
  \BibitemOpen
  \bibfield  {author} {\bibinfo {author} {\bibfnamefont {G.}~\bibnamefont
  {D'Amico}}, \bibinfo {author} {\bibfnamefont {P.}~\bibnamefont {Panci}}, \
  and\ \bibinfo {author} {\bibfnamefont {A.}~\bibnamefont {Strumia}},\
  }\href@noop {} {\  (\bibinfo {year} {2018})},\ \Eprint
  {http://arxiv.org/abs/1803.03629} {arXiv:1803.03629 [astro-ph.CO]}
  \BibitemShut {NoStop}%
\bibitem [{\citenamefont {Kang}(2018)}]{Kang:2018qhi}%
  \BibitemOpen
  \bibfield  {author} {\bibinfo {author} {\bibfnamefont {Z.}~\bibnamefont
  {Kang}},\ }\href@noop {} {\  (\bibinfo {year} {2018})},\ \Eprint
  {http://arxiv.org/abs/1803.04928} {arXiv:1803.04928 [hep-ph]} \BibitemShut
  {NoStop}%
\bibitem [{\citenamefont {Yang}\ \emph {et~al.}(2018)\citenamefont {Yang},
  \citenamefont {Huang},\ and\ \citenamefont {Feng}}]{Yang:2018gjd}%
  \BibitemOpen
  \bibfield  {author} {\bibinfo {author} {\bibfnamefont {Y.}~\bibnamefont
  {Yang}}, \bibinfo {author} {\bibfnamefont {X.}~\bibnamefont {Huang}}, \ and\
  \bibinfo {author} {\bibfnamefont {L.}~\bibnamefont {Feng}},\ }\href@noop {}
  {\  (\bibinfo {year} {2018})},\ \Eprint {http://arxiv.org/abs/1803.05803}
  {arXiv:1803.05803 [astro-ph.CO]} \BibitemShut {NoStop}%
\bibitem [{\citenamefont {Pospelov}\ \emph {et~al.}(2018)\citenamefont
  {Pospelov}, \citenamefont {Pradler}, \citenamefont {Ruderman},\ and\
  \citenamefont {Urbano}}]{Pospelov:2018kdh}%
  \BibitemOpen
  \bibfield  {author} {\bibinfo {author} {\bibfnamefont {M.}~\bibnamefont
  {Pospelov}}, \bibinfo {author} {\bibfnamefont {J.}~\bibnamefont {Pradler}},
  \bibinfo {author} {\bibfnamefont {J.~T.}\ \bibnamefont {Ruderman}}, \ and\
  \bibinfo {author} {\bibfnamefont {A.}~\bibnamefont {Urbano}},\ }\href@noop {}
  {\  (\bibinfo {year} {2018})},\ \Eprint {http://arxiv.org/abs/1803.07048}
  {arXiv:1803.07048 [hep-ph]} \BibitemShut {NoStop}%
\bibitem [{\citenamefont {Safarzadeh}\ \emph {et~al.}(2018)\citenamefont
  {Safarzadeh}, \citenamefont {Scannapieco},\ and\ \citenamefont
  {Babul}}]{Safarzadeh:2018hhg}%
  \BibitemOpen
  \bibfield  {author} {\bibinfo {author} {\bibfnamefont {M.}~\bibnamefont
  {Safarzadeh}}, \bibinfo {author} {\bibfnamefont {E.}~\bibnamefont
  {Scannapieco}}, \ and\ \bibinfo {author} {\bibfnamefont {A.}~\bibnamefont
  {Babul}},\ }\href@noop {} {\  (\bibinfo {year} {2018})},\ \Eprint
  {http://arxiv.org/abs/1803.08039} {arXiv:1803.08039 [astro-ph.CO]}
  \BibitemShut {NoStop}%
\bibitem [{\citenamefont {Clark}\ \emph {et~al.}(2018)\citenamefont {Clark},
  \citenamefont {Dutta}, \citenamefont {Gao}, \citenamefont {Ma},\ and\
  \citenamefont {Strigari}}]{Clark:2018ghm}%
  \BibitemOpen
  \bibfield  {author} {\bibinfo {author} {\bibfnamefont {S.}~\bibnamefont
  {Clark}}, \bibinfo {author} {\bibfnamefont {B.}~\bibnamefont {Dutta}},
  \bibinfo {author} {\bibfnamefont {Y.}~\bibnamefont {Gao}}, \bibinfo {author}
  {\bibfnamefont {Y.-Z.}\ \bibnamefont {Ma}}, \ and\ \bibinfo {author}
  {\bibfnamefont {L.~E.}\ \bibnamefont {Strigari}},\ }\href@noop {} {\
  (\bibinfo {year} {2018})},\ \Eprint {http://arxiv.org/abs/1803.09390}
  {arXiv:1803.09390 [astro-ph.HE]} \BibitemShut {NoStop}%
\bibitem [{\citenamefont {Cheung}\ \emph {et~al.}(2018)\citenamefont {Cheung},
  \citenamefont {Kuo}, \citenamefont {Ng},\ and\ \citenamefont
  {Tsai}}]{Cheung:2018vww}%
  \BibitemOpen
  \bibfield  {author} {\bibinfo {author} {\bibfnamefont {K.}~\bibnamefont
  {Cheung}}, \bibinfo {author} {\bibfnamefont {J.-L.}\ \bibnamefont {Kuo}},
  \bibinfo {author} {\bibfnamefont {K.-W.}\ \bibnamefont {Ng}}, \ and\ \bibinfo
  {author} {\bibfnamefont {Y.-L.~S.}\ \bibnamefont {Tsai}},\ }\href@noop {} {\
  (\bibinfo {year} {2018})},\ \Eprint {http://arxiv.org/abs/1803.09398}
  {arXiv:1803.09398 [astro-ph.CO]} \BibitemShut {NoStop}%
\bibitem [{\citenamefont {Slatyer}\ and\ \citenamefont
  {Wu}(2018)}]{Slatyer:2018aqg}%
  \BibitemOpen
  \bibfield  {author} {\bibinfo {author} {\bibfnamefont {T.~R.}\ \bibnamefont
  {Slatyer}}\ and\ \bibinfo {author} {\bibfnamefont {C.-L.}\ \bibnamefont
  {Wu}},\ }\href@noop {} {\  (\bibinfo {year} {2018})},\ \Eprint
  {http://arxiv.org/abs/1803.09734} {arXiv:1803.09734 [astro-ph.CO]}
  \BibitemShut {NoStop}%
\bibitem [{\citenamefont {Liu}\ and\ \citenamefont
  {Slatyer}(2018)}]{Liu:2018uzy}%
  \BibitemOpen
  \bibfield  {author} {\bibinfo {author} {\bibfnamefont {H.}~\bibnamefont
  {Liu}}\ and\ \bibinfo {author} {\bibfnamefont {T.~R.}\ \bibnamefont
  {Slatyer}},\ }\href@noop {} {\  (\bibinfo {year} {2018})},\ \Eprint
  {http://arxiv.org/abs/1803.09739} {arXiv:1803.09739 [astro-ph.CO]}
  \BibitemShut {NoStop}%
\bibitem [{\citenamefont {Falkowski}\ and\ \citenamefont
  {Petraki}(2018)}]{Falkowski:2018qdj}%
  \BibitemOpen
  \bibfield  {author} {\bibinfo {author} {\bibfnamefont {A.}~\bibnamefont
  {Falkowski}}\ and\ \bibinfo {author} {\bibfnamefont {K.}~\bibnamefont
  {Petraki}},\ }\href@noop {} {\  (\bibinfo {year} {2018})},\ \Eprint
  {http://arxiv.org/abs/1803.10096} {arXiv:1803.10096 [hep-ph]} \BibitemShut
  {NoStop}%
\bibitem [{\citenamefont {Prinz}\ \emph {et~al.}(1998)\citenamefont {Prinz}
  \emph {et~al.}}]{Prinz:1998ua}%
  \BibitemOpen
  \bibfield  {author} {\bibinfo {author} {\bibfnamefont {A.~A.}\ \bibnamefont
  {Prinz}} \emph {et~al.},\ }\href {\doibase 10.1103/PhysRevLett.81.1175}
  {\bibfield  {journal} {\bibinfo  {journal} {Phys. Rev. Lett.}\ }\textbf
  {\bibinfo {volume} {81}},\ \bibinfo {pages} {1175} (\bibinfo {year}
  {1998})},\ \Eprint {http://arxiv.org/abs/hep-ex/9804008}
  {arXiv:hep-ex/9804008 [hep-ex]} \BibitemShut {NoStop}%
\bibitem [{\citenamefont {Vogel}\ and\ \citenamefont
  {Redondo}(2014)}]{Vogel:2013raa}%
  \BibitemOpen
  \bibfield  {author} {\bibinfo {author} {\bibfnamefont {H.}~\bibnamefont
  {Vogel}}\ and\ \bibinfo {author} {\bibfnamefont {J.}~\bibnamefont
  {Redondo}},\ }\href {\doibase 10.1088/1475-7516/2014/02/029} {\bibfield
  {journal} {\bibinfo  {journal} {JCAP}\ }\textbf {\bibinfo {volume} {1402}},\
  \bibinfo {pages} {029} (\bibinfo {year} {2014})},\ \Eprint
  {http://arxiv.org/abs/1311.2600} {arXiv:1311.2600 [hep-ph]} \BibitemShut
  {NoStop}%
\bibitem [{\citenamefont {Chang}\ \emph {et~al.}(2018)\citenamefont {Chang},
  \citenamefont {Essig},\ and\ \citenamefont {McDermott}}]{Chang:2018rso}%
  \BibitemOpen
  \bibfield  {author} {\bibinfo {author} {\bibfnamefont {J.~H.}\ \bibnamefont
  {Chang}}, \bibinfo {author} {\bibfnamefont {R.}~\bibnamefont {Essig}}, \ and\
  \bibinfo {author} {\bibfnamefont {S.~D.}\ \bibnamefont {McDermott}},\
  }\href@noop {} {\  (\bibinfo {year} {2018})},\ \Eprint
  {http://arxiv.org/abs/1803.00993} {arXiv:1803.00993 [hep-ph]} \BibitemShut
  {NoStop}%
\bibitem [{\citenamefont {McDermott}\ \emph {et~al.}(2011)\citenamefont
  {McDermott}, \citenamefont {Yu},\ and\ \citenamefont
  {Zurek}}]{McDermott:2010pa}%
  \BibitemOpen
  \bibfield  {author} {\bibinfo {author} {\bibfnamefont {S.~D.}\ \bibnamefont
  {McDermott}}, \bibinfo {author} {\bibfnamefont {H.-B.}\ \bibnamefont {Yu}}, \
  and\ \bibinfo {author} {\bibfnamefont {K.~M.}\ \bibnamefont {Zurek}},\ }\href
  {\doibase 10.1103/PhysRevD.83.063509} {\bibfield  {journal} {\bibinfo
  {journal} {Phys. Rev.}\ }\textbf {\bibinfo {volume} {D83}},\ \bibinfo {pages}
  {063509} (\bibinfo {year} {2011})},\ \Eprint {http://arxiv.org/abs/1011.2907}
  {arXiv:1011.2907 [hep-ph]} \BibitemShut {NoStop}%
\bibitem [{\citenamefont {Fennelly}\ and\ \citenamefont
  {Torr}(1992)}]{FENNELLY1992321}%
  \BibitemOpen
  \bibfield  {author} {\bibinfo {author} {\bibfnamefont {J.}~\bibnamefont
  {Fennelly}}\ and\ \bibinfo {author} {\bibfnamefont {D.}~\bibnamefont
  {Torr}},\ }\href {\doibase https://doi.org/10.1016/0092-640X(92)90004-2}
  {\bibfield  {journal} {\bibinfo  {journal} {Atomic Data and Nuclear Data
  Tables}\ }\textbf {\bibinfo {volume} {51}},\ \bibinfo {pages} {321 }
  (\bibinfo {year} {1992})}\BibitemShut {NoStop}%
\bibitem [{\citenamefont {{Popov}}(1999)}]{1999TJPh...23..943P}%
  \BibitemOpen
  \bibfield  {author} {\bibinfo {author} {\bibfnamefont {V.}~\bibnamefont
  {{Popov}}},\ }\href@noop {} {\bibfield  {journal} {\bibinfo  {journal}
  {Turkish Journal of Physics}\ }\textbf {\bibinfo {volume} {23}},\ \bibinfo
  {pages} {943} (\bibinfo {year} {1999})}\BibitemShut {NoStop}%
\bibitem [{\citenamefont
  {Karshenboim}(2010{\natexlab{a}})}]{Karshenboim:2010ck}%
  \BibitemOpen
  \bibfield  {author} {\bibinfo {author} {\bibfnamefont {S.~G.}\ \bibnamefont
  {Karshenboim}},\ }\href {\doibase 10.1103/PhysRevD.82.073003} {\bibfield
  {journal} {\bibinfo  {journal} {Phys. Rev.}\ }\textbf {\bibinfo {volume}
  {D82}},\ \bibinfo {pages} {073003} (\bibinfo {year} {2010}{\natexlab{a}})},\
  \Eprint {http://arxiv.org/abs/1005.4872} {arXiv:1005.4872 [hep-ph]}
  \BibitemShut {NoStop}%
\bibitem [{\citenamefont
  {Karshenboim}(2010{\natexlab{b}})}]{Karshenboim:2010cg}%
  \BibitemOpen
  \bibfield  {author} {\bibinfo {author} {\bibfnamefont {S.~G.}\ \bibnamefont
  {Karshenboim}},\ }\href {\doibase 10.1103/PhysRevLett.104.220406} {\bibfield
  {journal} {\bibinfo  {journal} {Phys. Rev. Lett.}\ }\textbf {\bibinfo
  {volume} {104}},\ \bibinfo {pages} {220406} (\bibinfo {year}
  {2010}{\natexlab{b}})},\ \Eprint {http://arxiv.org/abs/1005.4859}
  {arXiv:1005.4859 [hep-ph]} \BibitemShut {NoStop}%
\bibitem [{\citenamefont {Aguilar-Arevalo}\ \emph {et~al.}(2017)\citenamefont
  {Aguilar-Arevalo} \emph {et~al.}}]{Aguilar-Arevalo:2016zop}%
  \BibitemOpen
  \bibfield  {author} {\bibinfo {author} {\bibfnamefont {A.}~\bibnamefont
  {Aguilar-Arevalo}} \emph {et~al.} (\bibinfo {collaboration} {DAMIC}),\ }\href
  {\doibase 10.1103/PhysRevLett.118.141803} {\bibfield  {journal} {\bibinfo
  {journal} {Phys. Rev. Lett.}\ }\textbf {\bibinfo {volume} {118}},\ \bibinfo
  {pages} {141803} (\bibinfo {year} {2017})},\ \Eprint
  {http://arxiv.org/abs/1611.03066} {arXiv:1611.03066 [astro-ph.CO]}
  \BibitemShut {NoStop}%
\bibitem [{\citenamefont {An}\ \emph {et~al.}(2013)\citenamefont {An},
  \citenamefont {Pospelov},\ and\ \citenamefont {Pradler}}]{An:2013yfc}%
  \BibitemOpen
  \bibfield  {author} {\bibinfo {author} {\bibfnamefont {H.}~\bibnamefont
  {An}}, \bibinfo {author} {\bibfnamefont {M.}~\bibnamefont {Pospelov}}, \ and\
  \bibinfo {author} {\bibfnamefont {J.}~\bibnamefont {Pradler}},\ }\href
  {\doibase 10.1016/j.physletb.2013.07.008} {\bibfield  {journal} {\bibinfo
  {journal} {Phys. Lett.}\ }\textbf {\bibinfo {volume} {B725}},\ \bibinfo
  {pages} {190} (\bibinfo {year} {2013})},\ \Eprint
  {http://arxiv.org/abs/1302.3884} {arXiv:1302.3884 [hep-ph]} \BibitemShut
  {NoStop}%
\bibitem [{\citenamefont {Redondo}\ and\ \citenamefont
  {Raffelt}(2013)}]{Redondo:2013lna}%
  \BibitemOpen
  \bibfield  {author} {\bibinfo {author} {\bibfnamefont {J.}~\bibnamefont
  {Redondo}}\ and\ \bibinfo {author} {\bibfnamefont {G.}~\bibnamefont
  {Raffelt}},\ }\href {\doibase 10.1088/1475-7516/2013/08/034} {\bibfield
  {journal} {\bibinfo  {journal} {JCAP}\ }\textbf {\bibinfo {volume} {1308}},\
  \bibinfo {pages} {034} (\bibinfo {year} {2013})},\ \Eprint
  {http://arxiv.org/abs/1305.2920} {arXiv:1305.2920 [hep-ph]} \BibitemShut
  {NoStop}%
\bibitem [{\citenamefont {Agnese}\ \emph {et~al.}(2018)\citenamefont {Agnese}
  \emph {et~al.}}]{Agnese:2018col}%
  \BibitemOpen
  \bibfield  {author} {\bibinfo {author} {\bibfnamefont {R.}~\bibnamefont
  {Agnese}} \emph {et~al.} (\bibinfo {collaboration} {SuperCDMS}),\ }\href@noop
  {} {\bibfield  {journal} {\bibinfo  {journal} {Submitted to: Phys. Rev.
  Lett.}\ } (\bibinfo {year} {2018})},\ \Eprint
  {http://arxiv.org/abs/1804.10697} {arXiv:1804.10697 [hep-ex]} \BibitemShut
  {NoStop}%
\end{thebibliography}%

\end{document}